\documentclass[review]{elsarticle}

\usepackage{lineno,hyperref}
\usepackage{array}
\usepackage{amsmath,amsfonts,amssymb}
\usepackage{multirow}
\usepackage{color, colortbl}
\usepackage{comment}
\usepackage{xcolor}

\graphicspath{{./Img/}}

\modulolinenumbers[5]

\journal{Medical Image Analysis}

\biboptions{authoryear}

\begin{document}

\begin{frontmatter}

\title{Multi-objective optimization determines when, which and how to fuse deep networks: an application to predict COVID-19 outcomes}

\author[campus,sapienza]{Valerio Guarrasi\corref{correspondingauthor}}
\cortext[correspondingauthor]{Corresponding author}
\ead{valerio.guarrasi@uniroma1.com}
\author[campus]{Paolo Soda}
\address[campus]{Unit of Computer Systems and Bioinformatics, Department of Engineering, University Campus Bio-Medico of Rome, Italy}
\address[sapienza]{Department of Computer, Control, and Management Engineering, Sapienza University of Rome, Italy}

\begin{abstract}

The COVID-19 pandemic has caused millions of cases and deaths and the AI-related scientific community, after being involved with detecting COVID-19 signs in medical images, has been now directing the efforts towards the development of methods that can predict the progression of the disease.
This task is multimodal by its very nature and, recently, baseline results achieved on the publicly available AIforCOVID dataset have shown that chest X-ray scans and clinical information are useful to identify patients at risk of severe outcomes.
While deep learning has shown superior performance in several medical fields, in most of the cases it considers unimodal data only.
In this respect, when, which and how to fuse the different modalities is an open challenge in multimodal deep learning.
To cope with these three questions here we present a novel approach optimizing the setup of a multimodal end-to-end model.
It exploits Pareto multi-objective optimization working with a performance metric and the diversity score of multiple candidate unimodal neural networks to be fused. 
We test our method on the AIforCOVID dataset, attaining state-of-the-art results, not only outperforming the baseline performance but also being robust to external validation.
Moreover, exploiting XAI algorithms we figure out a hierarchy among the modalities and we extract the features' intra-modality importance, enriching the trust on the predictions made by the model.

\end{abstract}


\end{frontmatter}


\section{Introduction} \label{sec:Introduction}

It has been two years since the severe acute respiratory syndrome COVID-19 has struck the world with a pandemic causing millions of cases and deaths.
Since the beginning, the scientific community has tried to contain the spread and the number of victims by studying the virus from multiple perspectives and looking for the best cures.

Among the various initiatives to combat the rising of the pandemic, researchers, practitioners and enterprises introduced new artificial intelligence (AI) methods and tools to process chest X-ray (CXR) and computed tomography (CT) examinations to replace or to supplement the reverse transcriptase-polymerase chain reaction (RT-PCR) tests.
Both medical imaging procedures are instrumental in helping radiologists determine the source of symptoms, stratify the disease severity, and establish the best treatment plan for the patient's specific needs.
On the other hand, CXR helps indicate abnormal formations of a large variety of chest diseases by using a very small amount of radiation.
Reticular alteration (up to 5 days from the symptoms' onset) and ground-glass opacity (following 5 days from the symptoms' onset) are the CXR most frequent lesions in COVID-19 positive patients.
The patients' consolidation gradually increases over time, striking mainly bilateral, peripheral, middle/lower locations.
On the other hand, CT delivers a much higher level of detail, offering a computerized 360-degree view of the lungs' structures.
However, by choosing this modality, we observed the lack of availability of machines' slots, the difficulty of moving bedridden patients, and long sanitation times.
On the contrary, X-ray equipment is much smaller, less complex, and with lower costs than CT scans.
For these reasons, in this work we focus on the use of CXR scans, that fit well with the COVID-19 crisis.

Most of the AI models for medical image-based applications, in particular those exploiting deep-learning (DL) models, consider only pixel-value data neglecting information available in other modalities.
However, interpreting imaging findings is multimodal by its very nature and, hence, AI needs to be able to process together data coming from various modalities~\cite{bib:baltruvsaitis2018multimodal}.
Indeed, while multimodal learning is a well-established field of study~\cite{bib:baltruvsaitis2018multimodal,bib:ngiam2011multimodal}, the potential of deep-learning shown processing unimodal data has recently motivated the rise of multimodal deep-learning (MDL), which aims to treat multimodal data by using deep network-based approaches.
This should be very useful to progress in understanding the world around us, biomedicine included~\cite{bib:ramachandram2017deep}.
In the specific context of the COVID-19 pandemic, MDL can support the patients' stratification by mining together images and clinical information~\cite{bib:soda2021aiforcovid}.

The literature agrees that the three main open questions in MDL are \textit{when}, \textit{which} and \textit{how} the modalities must be joined to exploit their potential~\cite{bib:stahlschmidt2022multimodal}.
Usually, MDL models are constructed by finding or combining the best model architecture for each modality, which are then combined by researchers and practitioners on the basis of the nature of the data, the task at hand and the networks' structure.
However, this handcrafted approach does not necessarily provide the best ensemble~\cite{bib:chen2020pathomic}.
To overcome these limitations, here we present a novel joint fusion technique that algorithmically finds the optimal set of deep architectures to be combined, whatever the modality they belong to, by maximizing both a classification evaluation metric and a measure of diversity among the learners.
Indeed, diversity is an estimate of how the networks interpret data differently~\cite{bib:kuncheva2003measures}.
Next, the method trains the resulting architecture in an end-to-end fashion to carry out the task at hand. 
Further to determine which models and modalities should be fused, this work investigates when to join the different modalities and how to embed them in powerful multimodal data representations.
We apply the method to stratify patients suffering from the COVID-19 syndrome by predicting the severity of the outcome using both clinical and CXR scans, showing that we achieve state-of-the-art results.
Furthermore, to improve trust and transparency of our MDL approach, we present a novel combination of Explainable Artificial Intelligence (XAI) techniques not only illustrating the reasoning behind the decisions taken by the model, but also showing the relative contribution of each modality in making the decision.

The manuscript is organized as follows.
Section~\ref{sec:Background} introduces the state-of-the-art of both multimodal fusion methods and DL applications of COVID-19 stratification.
Section~\ref{sec:Materials}, illustrates the dataset on which the experiments were performed.
Section~\ref{sec:Methods} describes the novel fusion method presented in this work and the XAI algorithms used for interpretation. 
Section~\ref{sec:Experimental Configuration} introduces the experimental configuration in the context of COVID-19 prognosis prediction, whilst the corresponding results and explanations are offered in section~\ref{sec:Results and Discussions}.
Finally, section~\ref{sec:Conclusions} provides concluding remarks.

\section{Background} \label{sec:Background}

With the spread of the epidemic, there has been a growth of interest in employing novel AI methods to fight the pandemic~\cite{bib:chamola2020comprehensive}. 
The vast literature ranges from the use of traditional statistical models to the more complex DL networks.
As it is out of the scope of this manuscript to mention the entire literature, here we give an overview of the state-of-the-art of DL applications predicting the disease's outcome, focusing on those that used supervised MDL models and exploited a joint fusion between medical imaging and clinical data.
Since the MDL method presented in this manuscript fits well with XAI algorithms, we also give an overview of this emerging field of study. 

\subsection{Multimodal Deep Learning} \label{sec:fusion}
Multimodal data is composed of data originated by different sources observing a phenomena.
The objective of MDL is to work with the data in a complementary manner to satisfy the undertaken learning task: instead of using manually designed or handcrafted modality-specific features, via DL we can automatically learn and extract an embedded representation for each modality.
Furthermore, here we focus on supervised multimodal fusion that joins information from different modalities for a classification or regression task. 
The literature agrees that understanding \textit{when}, \textit{which} and \textit{how} modalities should be fused is the main challenge in MDL~\cite{bib:baltruvsaitis2018multimodal,bib:stahlschmidt2022multimodal}. 

To answer the first question (\textit{when}), in the review~\cite{bib:baltruvsaitis2018multimodal} the authors divide multimodal fusion in model-agnostic and model-based methods.
The former are feature-based or decision-based, being also named as early and late fusion, respectively.
Early fusion consists in the integration of the raw multimodal data into a single feature vector which is used as input for the DL model.
At its simplest form, early fusion consists in concatenating multimodal features, which can bring to a redundancy of information.
Oppositely, late fusion consists in using multiple models trained on separate modalities, whose decisions are joint via an aggregation function or a meta-learner, aiming to augment the performance of each unimodal model~\cite{bib:wu2003fusing}.
Because model-agnostic fusions can suppress intra- or inter-modality interactions, in the last years researchers have mainly focused on model-based methods that allow intermediate fusion, also referred to as joint fusion, to occur directly inside the models.
DL models transform raw inputs into higher-level representations through the alternation of linear and non-linear operations.
With a MDL model, we obtain these representations for each modality which will be fused into a hidden layer creating a joint multimodal representation.
Thanks to the fact that with DL we can have an end-to-end training of all the multimodal representation components and the fusion component, the training phase essentially consists in learning hierarchical representations from raw data, which give rise to an optimized intermediate-level fusion representation~\cite{bib:bengio2013representation}.
This is why we can consider joint fusion an evolution of early fusion.
Furthermore, inside a joint model, to understand when the unimodal architectures should be joined and to find the fusion structure, the majority of the available work follows a meticulous handcrafted approach.
For instance, in~\cite{bib:neverova2015moddrop,bib:karpathy2014large} the authors adopted a single fusion layer from which they implement a gradual fusion strategy to automatically comprehend when to fuse the modalities.

Let us now focus on the second challenge, which aims to understand \textit{which} modalities should be fused and with which models.
With the increasing number of high-performing deep architectures, the selection of the one best suited for each modality is a hard process.
The choice of which modality to fuse is usually based on intuition and on a trial and error approach that, usually, first fuses similar modalities and then tries to integrate disparate modalities~\cite{bib:baltruvsaitis2018multimodal}.
Furthermore, prior investigations on the data, as well as feature selection techniques, should allow us to find redundant information along the modalities~\cite{bib:kumar2014feature}.
Shifting the attention to the choice of which neural networks should be fused, the analysis of the literature shows that the deep architectures are usually chosen by analyzing the single modalities~\cite{bib:soda2021aiforcovid,bib:chen2020pathomic}, and no optimal multimodal architecture selection is overtaken.

\begin{figure}[]
\centering
\caption{Multimodal joint-fusion methods: a) concatenation method~\cite{bib:nojavanasghari2016deep} b) multiplicative method~\cite{bib:zadeh2017tensor}.}
\includegraphics[width=\textwidth]{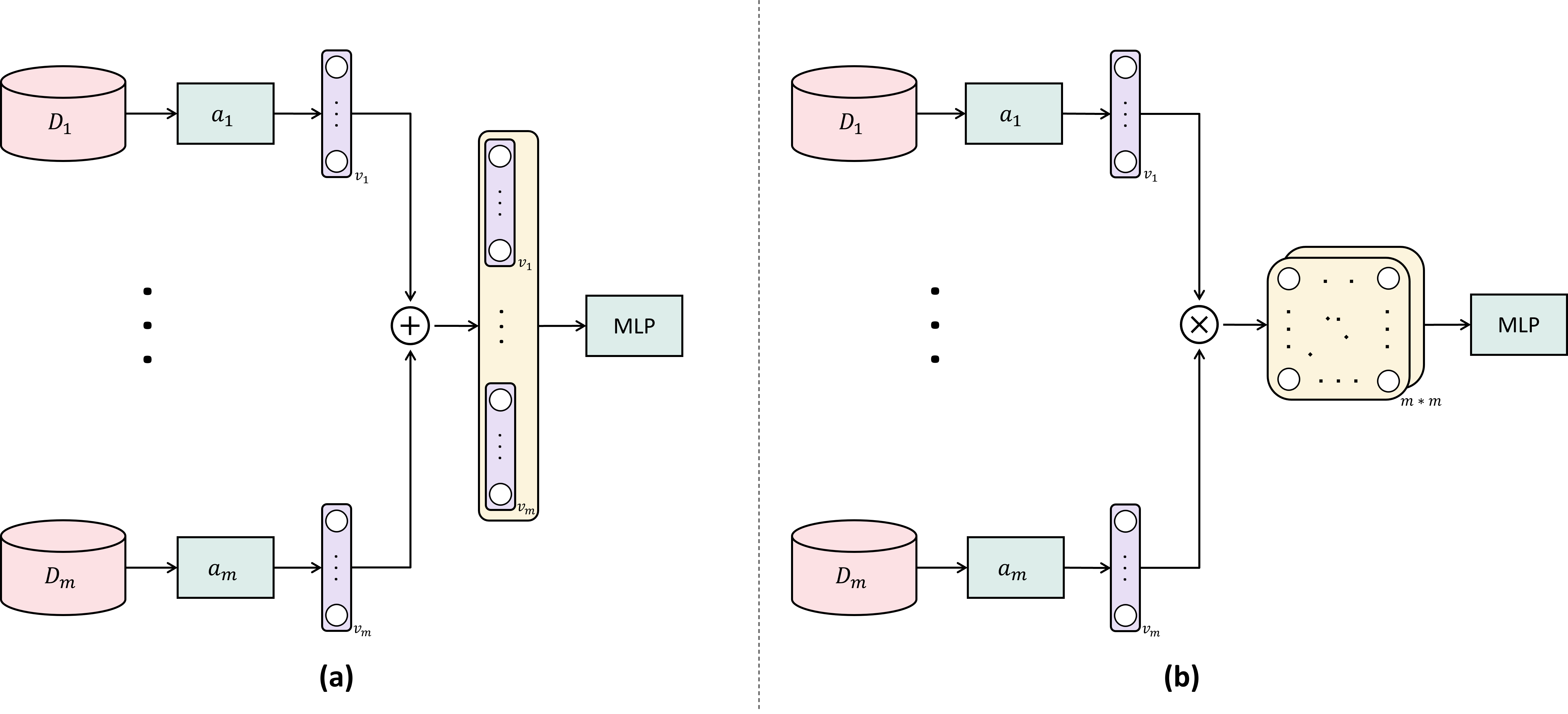}
\label{fig:joint}
\end{figure}

Focusing on the third question, in the literature there are various techniques studying \textit{how} to obtain an embedded representation that helps the unimodal models in the desired tasks.
In joint fusion, the state-of-the-art adopts two main methodologies~\cite{bib:nojavanasghari2016deep, bib:zadeh2017tensor}.
The first is a simple but effective way that concatenates the extracted vectors from a given layer of each unimodal model~\cite{bib:nojavanasghari2016deep}.
The next hidden layer combines the different modalities by computing
\begin{equation} \label{eq:concatenation}
f(\sum_{i=1}^{m} w_i^\intercal v_i)
\end{equation}
where $v_i$ and $w_i$ are the encoding and the weights of the model of the $i^{th}$ modality, $f$ is the activation function and $m$ stands for the number of modalities involved.
The second approach consists of a multiplicative method computing the outer product $\otimes$ (or Kronecker product) of all the $m$ modality-specific feature vectors, extrapolated from a given hidden layer, exploiting the intra- and inter-modality relations~\cite{bib:zadeh2017tensor}.
The layer output is:
\begin{equation} \label{eq:tensorfusion}
\bigotimes_{i=1}^{m}
\begin{bmatrix}
v_i \\
1
\end{bmatrix}
\end{equation}
where the 1 is appended to the encoding $v_i$ so that all unimodal and multimodal combinations interactions are represented.
A schematic visualization of both joint fusion methods is shown in figure~\ref{fig:joint}.

Many studies such as~\cite{bib:ramachandram2018bayesian,bib:kim2018hierarchy,bib:joze2020mmtm} dealt with determining an automatic methodology on how and when to fuse modalities, but none have focused on finding an optimized way for choosing which model architectures for each modality should be fused.
Therefore, we hereby propose a strategy to find the best combination of models of the different modalities. 
\subsection{DL and COVID-19} \label{DL_COVID}
The application of AI to the COVID-19 data analysis has been directed towards medical imaging and clinical data addressing two main issues: \emph{i}) the detection of patients suffering from COVID-19 pneumonia from those which are healthy or affected by different types of pneumonia and, \emph{ii}) the prediction of the outcome.

In the former case, a large number of publications have been produced, as reviewed in~\cite{bib:wynants2020prediction}.
Many of them leverage the open-source CXR COVID-19 data collection~\cite{bib:cohen2020covid}, or specific clinical datasets.
Moreover, there are also papers that use CT scans, as reviewed in~\cite{bib:ozsahin2020review}, or that use other imaging modalities~\cite{bib:wynants2020prediction}.
Instead, in the prediction scenario, few works have investigated the stratification of the disease to distinguish between patients with mild and severe outcomes~\cite{bib:soda2021aiforcovid,bib:signoroni2021bs,bib:zhu2020deep,bib:al2020classifier, bib:bai2020predicting,bib:ning2020open,bib:fang2021deep}.
In this respect, the rest of this subsection overviews such papers since outcome prediction is the focus of our work.

Considering the publications which work with CXR scans only, in~\cite{bib:signoroni2021bs} the authors designed an end-to-end convolutional neural network (CNN) architecture for predicting a multi-regional score conveying the degree of lung compromised in COVID-19 patients.

Focusing on the publications using clinical data only, in~\cite{bib:zhu2020deep} the authors presented a DL algorithm, whose architecture is composed of 6 fully connected (FC) dense layers with ReLU as activation function, which identifies the top clinical variable predictors and derived a risk stratification score system to help clinicians triage COVID-19 patients. 
Their model was trained on a private dataset with 181 patients and of 56 clinical variables, obtaining an AUC equal to 0.968.
Similarly, in~\cite{bib:al2020classifier} the authors built a neural network with one hidden layer to predict the status of recovered and death COVID-19 patients in South Korea.
They used seven different clinical variables and obtained an accuracy equal to 0.966 on a private dataset composed of 1308 patients.

As interpreting medical data is a multimodal process by its very nature, we now present the four works that use both clinical data and medical images, also noticing that three of them use CT images~\cite{bib:bai2020predicting,bib:ning2020open,bib:fang2021deep} and only one made use of CXR scans~\cite{bib:soda2021aiforcovid}.
In~\cite{bib:bai2020predicting} the authors introduced a DL-based method using 53 clinical features and quantitative CT data to detect mild patients with a potential malignant progression.
The authors use a multi-layer perceptron (MLP) on the clinical features creating an embedding that is concatenated with the flattened CT, which are then fed to a Long short-term memory (LSTM) network and a FC network, achieving an accuracy of 0.792 on a cohort of 199 patients belonging to a private dataset.
Next, in~\cite{bib:ning2020open} the authors use both CT images and 130 clinical features to discriminate between negative, mild and severe COVID-19 cases via a DL framework based on the VGG-16 and a 7-layer FC network. This network was trained on a public dataset containing 1521 patients, and it obtained an accuracy equal to 0.811.
Furthermore, the authors of~\cite{bib:fang2021deep} developed an early-warning system with DL techniques to predict COVID-19 malignant progression leveraging CT scans and the clinical data of the patients.
The authors use a 3D ResNet and MLP to encode the chest CT scans and the 61 clinical features, respectively, whose features are concatenated and fed into an LSTM and several FC networks to make the prediction, obtaining an accuracy equal to 0.877, on a private dataset with 1040 patients.

Let us now turn the attention to the only work that, to the best of our knowledge, uses CXR images and clinical for the COVID-19 prognosis~\cite{bib:soda2021aiforcovid}.
In this paper, the authors not only made publicly available a dataset designed for the mild and severe stratification task, but they also presented three different multimodal AI approaches that can be used as a baseline performance for future improvement.
In the first learning approach, also referred to as handcrafted approach (HC), the authors employed first order and texture features computed from the images, which are mined together with the clinical data feeding a supervised learner, such as Support Vector Machines and Random Forests.
The second approach is named hybrid (HYB) and it mixes automatically extracted features computed by a pre-trained CNN with the clinical data, which are then fed to a classifier as before.
The last approach is an end-to-end approach (ETE) exploiting together the clinical data and the raw CXR by using a multi-input CNN.
The best results are obtained by the ETE approach, whose results are reported and discussed in section~\ref{sec:Results and Discussions}.

This analysis of the literature shows that the development of AI-based models predicting the outcomes of COVID-19 patients still deserves further research efforts, since in the multimodal scenario only simple methods are employed with no interest in understanding \textit{when}, \textit{which} and \textit{how} the neural networks architectures should be used to find the appropriate fusion.
We decided to use~\cite{bib:soda2021aiforcovid} as our starting point, given the importance of their multimodal data, and we will show that our proposed MDL framework improves the performance available at the state-of-the-art.

\subsection{XAI}
The major disadvantage of neural network approaches is their lack of interpretability~\cite{bib:arrieta2020explainable}.
It is hard to understand what the predictions rely on, and which modalities and features hold an important role~\cite{bib:joshi2021review}.
For this reason the number of XAI algorithms has grown in the last few years.
In general, an XAI algorithm is one that produces information to make a model's functioning clear or easy to understand.
The literature makes a clear distinction among models that are interpretable by design (transparent models), and those that can be explained by means of external XAI techniques (post-hoc explainability).
Given the complex structure of DL algorithms, post-hoc model-specific XAI models have been developed~\cite{bib:montavon2017explaining, bib:shrikumar2017learning, bib:hendricks2018women, bib:selvaraju2017grad, bib:sundararajan2017axiomatic}.
Focusing our attention to tabular or image data, i.e. the one used in this work, in the literature we found algorithms presenting feature relevance explanations specific for these modalities taken alone.
For example, according to~\cite{bib:arrieta2020explainable}, Deep Taylor decomposition~\cite{bib:montavon2017explaining} and DeepLIFT~\cite{bib:shrikumar2017learning} are the most used XAI models specific for multi-layer neural networks which work on tabular data, Equalizer~\cite{bib:hendricks2018women} and Grad-CAM~\cite{bib:selvaraju2017grad} are suited for CNNs working on image data, whereas Integrated Gradients should be used by both network typologies~\cite{bib:sundararajan2017axiomatic}.

The literature is well advanced in models which work on a single modality~\cite{bib:arrieta2020explainable} but it lacks research for MDL.
In this respect, the multimodal method that we present brings along the possibility to apply and combine any unimodal XAI algorithms in a way to understand the contribution that each modality and each data within a modality has during the prediction.

\section{Materials} \label{sec:Materials}

We performed the experiments on the AIforCOVID imaging archive~\cite{bib:soda2021aiforcovid} since it is the only publicly available multimodal dataset on COVID-19 stratification at the time this research was carried out.
It includes 820 CXR scans and clinical data collected from six different hospitals at the time of hospitalization of symptomatic patients positive to the SARS-CoV-2 infection at the TR-PCR test.
The 34 clinical available parameters are listed in \tablename~\ref{tab:clinicaldata}, and for further details the interested readers can refer to~\cite{bib:soda2021aiforcovid}.

On the basis of the clinical outcome, each patient was assigned to the mild or severe group, where the mild group consists of the patients that were sent back to domiciliary isolation or were hospitalized without any ventilatory support, whilst the severe group is composed of patients who required non-invasive ventilation support, intensive care unit or patients who deceased.

Recently, a second release of the dataset was made available and it is composed of other 283 patients, whose data were collected in other two new centers.
Here, we use these data as an external validation set in a way to have further proof that the proposed framework is robust on never seen data.

\section{Methods} \label{sec:Methods}

\subsection{Multimodal Learning}

\begin{figure}[]
\centering
\caption{Multimodal joint-late method: for each $m$ modalities, $n$ models are trained to find the optimal combination $\Gamma^*$, whose classification vector is passed to a FC neural network which will output the desired classification out of the $c$ classes.}
\includegraphics[width=\textwidth]{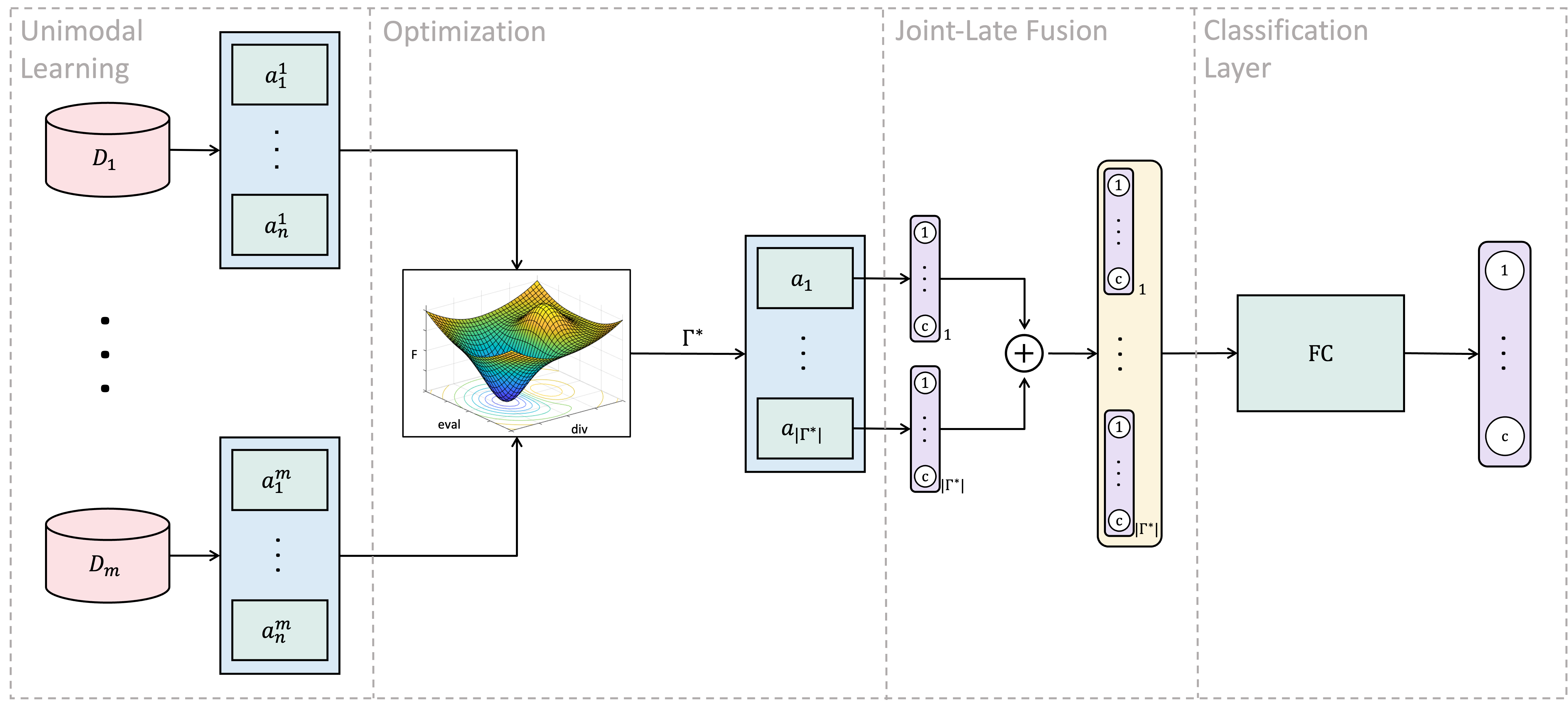}
\label{fig:model}
\end{figure}

To predict the severity outcome of COVID-19 patients exploiting both the pixel information of their CXR scan and their clinical information, we propose a novel supervised end-to-end joint fusion method, which maximizes the advantages brought by all the given modalities.
It first looks for the best combination of models, which are then joint and trained to carry out the given classification task.
This is visually summarized in \figurename~\ref{fig:model}, which is organized in four main blocks that are now detailed, one per paragraph.

\paragraph{Unimodal Learning}
The proposed method can be applied to an arbitrary number of $m$ different modalities and $n$ different neural network architectures.

Let us first introduce the application matrix $\Theta \in \text{Mat}_{m \times n}$, where each element is defined as follows
\begin{equation} \label{eq:matrix}
(\Theta)_{ij} = \theta_{ij} = \left\{ \begin{array}{rcl}
1 & \text{if the } j^{th} \text{model processes data of the } i^{th} \text{modality} \\
0 & \text{if the } j^{th} \text{model does not process data of the } i^{th} \text{modality}
\end{array}\right.
\end{equation}
Hence, $\theta_{ij}$ represents the use of the $j^{th}$ model for the $i^{th}$ modality, with $i \in \{1,\ldots,m\}$ and $j \in \{1,\ldots,n\}$.
Furthermore, let us denote with $a_j$ any neural network architecture we would like to use; on this basis, when $\theta_{ij}=1$, $a_j^i$ denotes that the architecture $a_j$ is trained on data from the $i^{th}$ modality.
The set of networks that will be combined is denoted by 
\begin{equation} \label{eq:combinantion}
\Gamma_I = \{a_{j}^{j} | \theta_{ij} = 1 , \forall (i,j) \in I\}
\end{equation}
where $I$ contains a subset of all possible pairs of models and modalities:
\begin{equation} \label{eq:I}
I \subseteq \{1,\ldots,m\} \times \{1,\ldots,n\}
\end{equation}
Consequently, the number of different configurations for $\Gamma_I$ is equal to $\sum_{h=2}^{s} \binom{s}{h}$, where $s=\sum_{i=1}^{m}\sum_{j=1}^{n} \theta_{ij}$.
Furthermore, model architectures being part of $\Gamma_I$ depend on the corresponding $\Theta$ matrix, whilst the choice of the number of models $n$ and which models to include should consider the constraints of the application, such as any limit in computational resources.

It is worth noting that the framework now introduced goes beyond the state-of-the-art of multimodal joint networks, which usually use only one model for each modality~\cite{bib:chen2020pathomic}.
On the contrary, we consider the chance of having several models per modality, a situation that researchers and practitioners often meet in practice.
In these cases, usually they first heuristically determine the best model per modality in a pool of many, i.e. $\Gamma_I$, and they combine such``best'' models in a multimodal learner~\cite{bib:arrieta2020explainable}.
However, this approach neglects any possible interaction between models and modalities, so that the combination of best unimodal models does not necessarily reflect the best multimodal architecture.
Furthermore, the general assumption that a multimodal model should be composed of a single model per modality is also a too much straightforward simplification.

\paragraph{Optimization} \label{sec:optimization}

To go beyond the naive use of $\Gamma_I$, we deem that it should be more useful to work with all the modalities and all the models together.
The first step is to detect which is the subset of models to be combined to get the best multimodal architecture providing the best multimodal data representation.
This subset is determined by solving a multi-objective maximization problem that leverages on two scores computed on a validation set for each $\Gamma_I$~\cite{bib:guarrasi2022pareto,bib:guarrasi2021multi}.
The first score is any measure derived from the confusion matrix, shortly referred as $\text{eval}(\Gamma_I)$, whilst the second is a measure of the diversity between the unimodal learning models, denoted as $\text{div}(\Gamma_I)$.
To find optimal combination of models from $\Gamma_I$ exploiting at best all the modalities, denoted by $\Gamma^*$, the algorithm works as follows. 
Using a cross-validation approach, we first apply each model $a_j^i$, that was trained on the corresponding unimodal dataset $D_i$, to a validation set. 
Then, we compute the average of both $\text{eval}(\Gamma_I)$ and $\text{div}(\Gamma_I)$ among the cross-validation runs for all the possible $\Gamma_I$s, finally finding the combination $\Gamma^*$ which maximizes both metrics.
This means that we look for the $\Gamma_I$ that, on the one side, returns the best classification performance and, on the other side, reduces the incidence and effect of coincident errors among its members, thus considering possible relationships between models and modalities.

It is worth noting that such a process corresponds to solving a multi-objective optimization problem.
In fact, given the pair $P=(F,C)$, where $F=\{f_1(x), \ldots f_a(x)\}$ is the set of $a$ objective functions that we want to maximize and $C=\{c_1(x) \leq 0,\ldots,c_b(x) \leq 0\}$ are the $b$ constraints, where if $x\in C$ it is said to be an admissible point for $P$.
Given two admissible points $x_1$ and $x_2$ for $P$, $x_1 \succ x_2$ means that $x_1$ dominates $x_2$.
This means that $f_{\alpha}(x_1)\leq f_{\alpha}(x_2)$ $\forall \alpha \in \{1,\ldots, a \}$, and $f_{\alpha}(x_1)< f_{\alpha}(x_2)$ for at least one $\alpha \in \{1,\ldots, a\}$.
A Pareto optimum $x^*$ is an admissible point for $P$ if $x^*\succ x$ for any other admissible points $x$ for $P$. i.e. it tries to minimize all functions in $F$ under the constraints $C$, solving a multi-objective minimization problem over $P$.
Based on~\cite{bib:john2014extremum}, the point $x^*$ is a Pareto optimum if there exist $\lambda$ and $\mu$ satisfying the following system:
\begin{equation}\begin{cases} \sum_{\alpha=1}^a \lambda_{\alpha} \nabla f_{\alpha}(x^*) + \sum_{\beta=1}^b \mu_{\beta} \nabla c_{\beta}(x^*)=0 \\ \sum_{\beta=1}^b \mu_{\beta} c_{\beta}(x^*)=0 \\ \lambda_{\alpha} \geq 0 \;\; \forall \alpha=1,\ldots a \\ \mu_{\beta} \geq 0 \;\; \forall \beta=1,\ldots b \\ (\boldsymbol{\lambda},\boldsymbol{\mu}) \neq (0,0) \end{cases}\end{equation}

In our framework, $\Gamma^*$ is the Pareto Optimum maximizing both two scores, related to the confusion matrix and to the diversity, respectively.
This implies that $\text{eval}(\Gamma_I)$ and $\text{div}(\Gamma_I)$ are the two objective functions ($a=2$).
Thus, $\Gamma^*$ is a Pareto optimum if the quintuple $(\lambda_1^*,\lambda_2^*, \mu_1^*,\mu_2^*, \Gamma^*)$ satisfies the following system: 
\begin{equation}\begin{cases} \lambda_1(-\dot{\text{eval}}(\Gamma_I)) + \lambda_2(-\dot{\text{div}}(\Gamma_I)) + \mu_1 - \mu_2 = 0 \\ \lambda_1\geq0, \lambda_2\geq0, \mu_1\geq0, \mu_2\geq0 \end{cases}\end{equation}
where $\dot{\text{eval}}(\Gamma_I)$ and $\dot{\text{div}}(\Gamma_I)$ denotes the first derivative of such quantities.
Observing that both $\text{eval}(\Gamma_I)$ and $\text{div}(\Gamma_I)$ range in $[0,1]$, and under the hypothesis that $\text{eval}(\Gamma_I)=1$ in case of perfect outcome prediction \footnote{Straightforwardly, if $\text{eval}(\Gamma_I)$ is an error-related measure, i.e. $\text{eval}(\Gamma_I)=0$ in case of perfect outcome prediction, equation~\ref{eq:objective} can be rewritten as $\text{eval}(\Gamma_I)^2 + (\text{div}(\Gamma_I)-1)^2$.}
we can write the following minimization function:
\begin{equation}
\label{eq:objective}
(\text{eval}(\Gamma_I)-1)^2 + (\text{div}(\Gamma_I)-1)^2
\end{equation}
Following~\cite{bib:kuhn2014nonlinear}, we can solve this system:
\begin{equation}\begin{cases} 2(\text{eval}(\Gamma_I)-1)\dot{\text{\text{eval}}}(\Gamma_I)+2(\text{div}(\Gamma_I)-1)\dot{\text{div}}(\Gamma_I)+\mu_1+\mu_2=0 \\ \mu_1\geq0,\;\; \mu_2\geq0 \end{cases}\end{equation}
Going forward this can be rewritten as: 
\begin{equation}\begin{cases} 2(1-\text{eval}(\Gamma_I))(-\dot{\text{eval}}(\Gamma_I))+2(1-\text{div}(\Gamma_I))(-\dot{\text{div}}(\Gamma_I))+\mu_1+\mu_2=0 \\ \mu_1\geq0,\;\; \mu_2\geq0 \end{cases}\end{equation}
$(\lambda_1^*,\lambda_2^*,\mu_1^*,\mu_2^*,\Gamma^*)$ satisfies the system and $\Gamma^*$ is a Pareto optimum for our optimization problem, when $\lambda_1^*=2(1-\text{eval}(\Gamma_I))$ and $\lambda_2^*=2(1-\text{div}(\Gamma_I))$, where $\lambda_1^*\geq0$ and $\lambda_2^*\geq0$ since $0\leq \text{eval}(\Gamma_I)\leq1$ and $0\leq div(\Gamma_I)\leq1$).

Moreover, with this optimization approach, if a modality is useless for the given classification task, all the networks working on that modality will be discarded.

\paragraph{Joint-Late Fusion} \label{sec:jointlate}
So far we presented an approach to find the optimal multimodal combination model, i.e. \textit{which} modalities and related models should be included into the multimodal architecture, but we have not defined \textit{how} and \textit{when} to actually join the models $a_j^i$ in $\Gamma_I$.
To simplify the notation, we now omit the apex $i$, denoting the learners included in $\Gamma_I$ as $a_j$, with $j = 1, \dots, |\Gamma_I|$.
Each of these models provides a classification vector $\mathbf{y}_j = [y_{j1}, \dots, y_{jp}, \dots , y_{jc}]^T$, where $c$ is the number of classes and $p$ is the generic $p^{th}$ output neuron.
Each entry $y_{jp}$ is defined using the parameterized softmax function:
\begin{equation}
 y_{jp} = \text{softmax}_k(z_p^{(j)})= 
 \frac{e^{k z_p^{(j)}} }{\sum_{t=1}^{c} e^{k z_t^{(j)}} }
\end{equation}
where $k > 0$, and $z_p^{(j)}$ denotes the $p^{th}$ value of the output layer neuron internal activity for the network $a_j$; similar definition holds for $z_t^{(j)}$.
Note that for $k=1$ each element represents an estimate of the classification posterior probability for each class of each model.

On these premises, to join the models we define a \textit{soft} shared representation given by $[\mathbf{y}_1^T, \dots, \mathbf{y}_{|\Gamma_I|}^T]^T$, which corresponds to concatenate the classification vectors, an approach similar to concatenating the hidden layers as in~\cite{bib:nojavanasghari2016deep}.
Hence, this shared representation vector is composed of $c \cdot |\Gamma_I|$ elements.
Furthermore, if $k \rightarrow \infty$, $\mathbf{y}_j$ becomes a binary vector with only one element equal to one, allowing us to get a \textit{crisp} shared representation containing the binary outputs, which provides a direct interpretation of the embedded layer.

\paragraph{Classification Layer}
This shared representation, being either soft or crisp, is then passed to a FC neural network, with $c$ neurons in the last layer, that performs the intermediate fusion.
Note that here we exploit the advantages of both joint and late fusion, since the classifications of the sub-networks are aggregated via an end-to-end manner using the back-propagation during the training process that minimizes the overall loss function. 
Indeed, it avoids the combination of features belonging to different spaces, while it combines the deep neural networks into a common classification space, and it is shortly referred to as to \textit{joint-late} fusion in the following. 

\subsection{XAI}

The nature of the joint-late fusion approach, makes interpretability simple and effective.
First, the weights $w$ coming out of the embedded classification vector connected to the first layer of the FC network, provides useful information because it reveals the importance of each class for each model and, consequently, the importance of each modality.
The higher the value of $w_i$ with $i=1,\cdots, n$, the higher is the contribution for the combined classification coming from $i^{th}$ model.
Furthermore, as more models per modality are allowed, we can also understand the models that contribute more in the final classification for each class by leaking at the weights proportion within each modality.

It is worth noting that any XAI algorithm, being model-agnostic or model-specific, can be applied to our MDL architecture, because it is composed of multiple models which can be interpreted one by one.
For instance, if one of the modalities uses tabular data and the chosen models are FC neural networks, any post-hoc model-specific XAI algorithm for multi-layer neural networks can be applied on the single models to understand the importance of each tabular variable~\cite{bib:montavon2017explaining, bib:shrikumar2017learning}.
Moreover, if another modality consists of images and the models used are CNNs, we can use a post-hoc model-specific XAI algorithm to deduce the contribution of each image's pixel to the classification~\cite{bib:hendricks2018women, bib:selvaraju2017grad}.
Going forward, there are also XAI algorithms that work on models regardless of the used modalities, that when applied to CNNs or FC networks, we understand the contributions for the classification of each image pixel and of each tabular variable, respectively~\cite{bib:sundararajan2017axiomatic}.
Since the model we propose is composed of multiple neural networks (both FC and CNNs as presented in section~\ref{sec:Experimental Configuration}), we can apply all the before-mentioned XAI algorithms on the corresponding network and sub-networks.

Given the fact in our framework we can have multiple models working with the same modality, we now introduce the concept of weighted XAI: it combines the outputs of the XAI algorithms specific to a modality with the weights coming from the embedded classification vector.
Since these weights make us understand the contribution of each model, we combine the different XAI importances via a weighted sum to obtain a single interpretation of the modality.
Hence for each feature $e_w$ the output of the weighted XAI algorithm importance is:
\begin{equation} \label{eq:weighted}
\mathbf{e}_w = \mathbf{e} \cdot \mathbf{w}
\end{equation}
where each contribution $\mathbf{e}_i$ extracted from the $n$ unimodal models is weighted with the corresponding layer weight $\mathbf{w}_i$ and summed together.
For example, if we are working on the image modality and we apply Grad-CAM~\cite{bib:selvaraju2017grad} to all the corresponding CNNs, we obtain a feature map for each model.
By computing the dot product between the maps and the weight vector, we get a single feature map explaining which pixels contributed more or less to that specific classification.
The same considerations hold for tabular data.

\section{Experimental Configuration} \label{sec:Experimental Configuration}

In this section we introduce the different experiments we performed.
Furthermore, to avoid any bias when comparing the baseline results of~\cite{bib:soda2021aiforcovid} that released the AIforCOVID dataset with the ones achieved by the learning framework presented here, we applied the same pre-processing procedure and validation approach, shortly described in subsections~\ref{subsec:pre-processing} and~\ref{subsec:validation}, respectively.

\subsection{Data Pre-processing} \label{subsec:pre-processing}
We use the 34 clinical descriptors available with the AIforCOVID repository, which are not direct indicators of the prognosis, that are listed in the appendix \tablename~\ref{tab:clinicaldata}.
Missing data were imputed using the mean and the mode for continuous and categorical variables, respectively.
Finally, to have the features all in the same range, a min-max scaler was applied along the variables.

In the case of CXR scans we extracted the segmentation mask of lungs, using a pre-trained U-Net~\cite{bib:ronneberger2015u} on two non-COVID-19 datasets~\cite{bib:shiraishi2000development, bib:jaeger2014two}. 
The mask was used to extrapolate the minimum squared bounding box containing both lungs, which is then resized to 224x224 and normalized with a min-max scaler bringing the pixel values between 0 and 1.
As already mentioned, the interested readers can refer to~\cite{bib:soda2021aiforcovid} for further details.

\subsection{Classification} 
As introduced in section~\ref{sec:Methods}, the first step to follow to obtain the optimal $\Gamma^*$ network is to train and evaluate different models using the data of the two modalities, i.e. CXR scans and clinical data.

For the image modality we worked with 30 different CNNs that come from 8 different main architectures: 
\begin{itemize}
 \item AlexNet~\cite{bib:krizhevsky2014one};
 \item VGG~\cite{bib:simonyan2014very}: VGG11, VGG11-BN, VGG13, VGG13-BN, VGG16, VGG16-BN, VGG19, VGG19-BN;
 \item ResNet~\cite{bib:he2016deep}: ResNet18, ResNet34, ResNet50, ResNet101, ResNet152, ResNeXt50, ResNeXt101, Wide-ResNet50-2, Wide-ResNet101-2;
 \item DenseNet~\cite{bib:huang2017densely}: DenseNet121, DenseNet169, DenseNet161, DenseNet201;
 \item GoogLeNet~\cite{bib:szegedy2015going};
 \item ShuffleNet~\cite{bib:ma2018shufflenet}: ShuffleNet-v2-x0-5, ShuffleNet-v2-x1-0, ShuffleNet-v2-x1-5, ShuffleNet-v2-x2-0;
 \item MobileNetV2~\cite{bib:sandler2018mobilenetv2};
 \item MNasNet~\cite{bib:tan2019mnasnet}: MNasNet0-5, MNasNet1-0;
\end{itemize}
In all the cases the weights were initialized using the values pre-trained on the ImageNet dataset~\cite{bib:deng2009imagenet}; we also changed the output layer dimension to 2 neurons, one for each class.

In the case of clinical information, which are tabular data, we adopted 4 MLPs that differ in terms of depth and wideness of the model.
We opted to use such architectures since these feedforward networks are able to learn a low-dimensional representation before being fused with the other modality~\cite{bib:glorot2010understanding}.
In particular, the models' hidden layers have the following organizations:
\begin{itemize}
 \item MLP-1: it has 3 hidden layers with 64, 64, 32 neurons respectively;
 \item MLP-2: it has 5 hidden layers with 64, 128, 128, 64, 32 neurons respectively;
 \item MLP-3: it has 7 hidden layers with 64, 128, 256, 256, 128, 64, 32 neurons respectively;
 \item MLP-4: it has 9 hidden layers with 64, 128, 256, 512, 512, 256, 128, 64, 32 neurons respectively;
\end{itemize}
A ReLU activation function is applied on all layers, since it learns several times faster than regular sigmoid activation functions~\cite{bib:glorot2011deep}, despite its non-differentiability at zero. 
Straightforwardly, the input layer and the output layers consist of 34 and 2 neurons, respectively.

Turning our attention to the matrix $\Theta$, it is composed of 24 columns ($n=24$ the number of architectures) and two rows ($m=2$ the number of modalities).

Furthermore, as for the two objective functions $\text{eval}(\Gamma_I)$ and $\text{div}(\Gamma_I)$ we adopt the accuracy ($\text{Acc}$) and the correlation coefficient ($\rho$), respectively.
Note that the accuracy is computed on the models part of the $\Gamma_I$s in late-fusion via the majority voting aggregation function.
We opted for late fusion optimization to prevent the training of all the $\sum_{h=2}^{s} \binom{s}{h}$ end-to-end ensemble models.
In this way, we only train $n$ single models, apply the optimization, and conclude with one final embedded training of $\Gamma^*$.

Majority voting consist in finding the most common classification $c$ between all the classifications $c_i$ extracted by the $|\Gamma_I|$ unimodal trained models
\begin{equation} \label{eq:majorityvoting}
c=Mo({c_i | 1 \leq i \leq |\Gamma_I| })
\end{equation}
where $Mo$ indicates the mode of a set and $|\Gamma_I|$ is the number of models in $\Gamma_I$.
The correlation coefficient $\rho$ instead between two classifiers is defined as:
\begin{equation}
\rho_{ij} = \frac{n^{11}n^{00}-n^{01}n^{10}}{\sqrt{(n^{11}+n^{10})(n^{01}+n^{00})(n^{11}+n^{01})(n^{10}+n^{00})}}
\end{equation}
where $n^{00}$ is the number of wrong classifications made by both models, $n^{11}$ is the number of correct classifications made by the models, $n^{10}$ and $n^{10}$ are the number of instances on which the classifications of the two models differ. 
Since $\rho$ is a pairwise measure, the overall diversity in $\Gamma_I$ is given by:
\begin{equation}
\label{eq:DiversityGammai}
 \rho(\Gamma_I) = \frac{2}{|\Gamma_I|(|\Gamma_I|-1)} \sum_{i=1}^{|\Gamma_I|-1} \sum_{j=i+1}^{|\Gamma_I|} \rho_{ij}
\end{equation}

Going beyond the single models, after determining $\Gamma^*$, its models are joint via the joint-late fusion method introduced in~\ref{sec:jointlate}.
As mentioned there, the classification space can be built adopting a soft and a crisp shared representation.
This permits us to investigate two learners, denoted as JLF-S and JLF-C, respectively.
We also investigate what happens by varying the final FC layer, lying after the shared representation.
To this end, we tested two different FC layers: the first does not have any hidden layers but directly connects the joint vector to the output classification layer using a single perceptron, and it is denoted using the ending 1.
The second consists of a single hidden layer of 4 neurons connected to the output classification layer, which is denoted using the ending 2.
The former permits us to investigate linear combinations among the classifications of the networks, whilst the latter investigates non-linear combinations.
In summary, we investigate four set-ups for the joint-late approach we present, i.e. JLF-S-1, JLF-S-2, JLF-C-1 and JLF-C-2.

\subsection{Training Configuration}

For both the single model and the multimodal scenarios, we adopt the training procedure described in~\cite{bib:soda2021aiforcovid}; so that any variation in performance is given only by the enhancements brought by the method and not by any other reason.
To prevent overfitting of the CNNs, we randomly applied the following image transformations with a probability equal to 0.3: horizontal or vertical shift ($-20 \leq$ pixels $\leq 20$), random zoom ($0.9 \leq$ factor $\leq 1.1$), vertical flip, random rotation ($-15^\circ \leq$ angle $\leq 15^\circ$), and elastic transform ($20 \leq \alpha \leq 40$, $\sigma=7$).
Both the CNNs and the MLPs were trained using the cross-entropy loss, regulated by an Adam optimizer with an initial learning rate of 0.001, which is scheduled to reduce by an order of magnitude every time the minimum validation loss doesn't change for 10 consecutive epochs.
To prevent overtraining and overfitting we fix the number of maximum epochs to 300, with an early stopping of 25 epochs on the validation loss.
We do not perform any preliminary optimization of the networks hyperparameters because in~\cite{arcuri2013parameter} the authors empirically reported that fine-tuning the learner parameters provides performance that do not statistically differ from those attained using the default values, a claim also confirmed by the experiments presented in~\cite{bib:soda2021aiforcovid} on the AIforCOVID dataset.
Furthermore, we do not pre-train the networks on another CXR dataset: indeed, although this would help the models learn modality-specific feature representations~\cite{rajaraman2020iteratively}, the results available on the AIforCOVID dataset show that this practice does not introduce any significant performance improvement~\cite{bib:soda2021aiforcovid}.

All the experiments were trained by using a batch size of 32 on a NVIDIA TESLA V100 GPU with 16 GB of memory, using PyTorch as the main DL coding library.

\subsection{Competitors} \label{subsec:Competitors}

To verify that the end-to-end process actually brings a performance enhancement, and to check that the joint network is not actually learning a late-fusion via FC layers using as input the classifications of the different networks, we froze the joint networks up to the joint classification vector and trained only the last FC classifier, creating a late fusion method.
We would expect that if there is an actual benefit in the joint end-to-end to fusion, these performances would be lower.
We will denote these experiments with: LF-S-1, LF-S-2, LF-C-1 and LF-C-2, where the S and the C encodings refer to the soft and the crisp embedded classification vectors and 1 and 2 refer to the type of FC layers employed, as before.
Finally, note that for all the experiments where joint training was performed (i.e. JLF-C-1, JLF-C-2, JLF-S-1, JLF-S-2, JF-C, JF-M), the weight initialization is performed by using the weights of the networks trained singularly.

We also compare the results of the joint-late fusion with those attained by the majority voting late fusion approach applied to the models included in $\Gamma^*$, denoted as LF-MV.
Since majority voting is the aggregation function used in the optimization phase, this comparison permits us to not only investigate if the joint-late training is beneficial, but also to compare our proposal against a late-fusion technique, addressing the question on \textit{when} the models should be fused.
We also tested the concatenation (denoted as JF-C) and the multiplication methods (denoted as JF-M) introduced in section~\ref{sec:fusion}: the comparison of these approaches against the JLF solutions lets us deepen \textit{how} the single models should be aggregated.
For the sake of completeness, we would note that the JLF approach intrinsically studies \textit{which} learners should be fused and, in the next section~\ref{sec:Results and Discussions} we also compare the results against the use of a single network.

Finally, as natural competitors, we also compare our results with those presented in the work introducing the AIforCOVID dataset~\cite{bib:soda2021aiforcovid}, i.e. the HC, HYB and ETE methods summarized in section~\ref{sec:Background}.

\subsection{Validation Approach} \label{subsec:validation}

The experiments were ran in 10-fold stratified cross-validation (CV), and leave-one-center-out cross-validation (LOCO), following the same experimental procedure described in~\cite{bib:soda2021aiforcovid}, thus ensuring a fair competitor between the approaches.
In each CV fold the proportion between the training, validation and testing sets is 70\%-20\%-10\%, respectively.

In LOCO validation we study how the models generalize to different data sources: indeed, in each fold the test set contains all the samples belonging to one center only, while the instances of the other six centers were assigned to the training and validation set.

With reference to~\cite{bib:soda2021aiforcovid}, here we add another layer of external validation (EV), ran on an external dataset.
To this goal, we exploited the second release of the AIforCOVID dataset, which contains data collected from 283 patients belonging to other two centers, as already mentioned in section~\ref{sec:Materials}.

\subsection{Performance metrics and statistical assessments}

The accuracy, the sensitivity and the specificity are the evaluation metrics used to assess the performance, as in~\cite{bib:soda2021aiforcovid}.

Furthermore, we apply the one-way ANOVA among the different groups of models and, to interpret the statistical significance, we used the pairwise Tukey test with a Bonferroni p-value correction at $\alpha=0.05$.
Henceforth, we refer to a statistically significant difference when this test is satisfied.

\section{Results and Discussions} \label{sec:Results and Discussions}

In this section we first show the results of the aforementioned experiments for the classification task, then we verify if the model is interpreting the data in the correct manner via the XAI methods introduced in section~\ref{sec:Methods}.

\subsection{Classification Task}

Before presenting the numeric results, let us report that the Pareto optimum $\Gamma^*$ is composed of three CNNs and one MLP, namely the GoogLeNet, the VGG13-BN, the ResNeXt50-32x4d, and the MLP-2.
It is worth noting that the selected models belong to different families suggesting that each extrapolates different information to satisfy the desired classification.
Moreover, we understand that the two modalities all give useful and distinct information for the prognosis task, since we have at least one model for each modality. 
Furthermore, we also ran the multi-objective optimization fusing the learners in each $\Gamma_I$ applying end-to-end training rather than late fusion to investigate if any differences exist\footnote{For computational restrictions during end-to-end training, we limited the number of possible architectures in $\Gamma_I$ for each modality to a maximum of 3.}.
The results reveal that both provided the same $\Gamma^*$, confirming that performing the optimization in late fusion rather in an end-to-end fashion is a viable alternative that strongly reduces the computation time since the former does not require any additional training.

The quantitative results achieved by all the learners presented in section~\ref{sec:Experimental Configuration} are shown in \tablename~\ref{tab:joint}, which shows the average accuracy (Acc), sensitivity (TPR) and specificity (TNR), with the respective standard deviations, for the CV, LOCO, and EV scenarios.
\begin{table}[]
\centering
\caption{Performance of all the learners. Note that results marked by an $^*$, i.e. those of HC, HYB and ETE are extracted from~\cite{bib:soda2021aiforcovid} and do not apply to the external dataset.}
\resizebox{\textwidth}{!}{
\begin{tabular}{l|l|ccc|ccc|ccc|}
\cline{2-11}
& \multirow{2}{*}{\textbf{Model}} & \multicolumn{3}{c|}{\textbf{CV}} & \multicolumn{3}{c|}{\textbf{LOCO}} & \multicolumn{3}{c|}{\textbf{EV}} \\
& & \textbf{Acc} & \textbf{TPR} & \textbf{TNR} & \textbf{Acc} & \textbf{TPR} & \textbf{TNR} & \textbf{Acc} & \textbf{TPR} & \textbf{TNR} \\ \hline
\multicolumn{1}{|l|}{\multirow{4}*{\rotatebox{90}{\textbf{Proposal}}}} & JLF-C-1 & 79.75$\pm$0.23 & 82.47$\pm$0.37 & 76.81$\pm$0.24 & 77.86$\pm$0.33 & 79.95$\pm$0.17 & 75.95$\pm$0.30 & 77.61$\pm$1.10 & 79.18$\pm$1.21 & 75.38$\pm$1.11 \\ \cline{2-11}
\multicolumn{1}{|l|}{} & JLF-C-2 & 79.63$\pm$0.24 & 80.99$\pm$0.17 & 78.90$\pm$0.23 & 77.30$\pm$0.32 & 79.06$\pm$0.22 & 75.88$\pm$0.17 & 76.82$\pm$1.22 & 79.50$\pm$1.26 & 74.23$\pm$1.23 \\ \cline{2-11}
\multicolumn{1}{|l|}{} & JLF-S-1 & 79.63$\pm$0.16 & 82.47$\pm$0.15 & 76.67$\pm$0.20 & 77.39$\pm$0.29 & 80.00$\pm$0.27 & 75.52$\pm$0.14 & 77.90$\pm$1.27 & 78.70$\pm$1.24 & 75.96$\pm$1.21 \\ \cline{2-11}
\multicolumn{1}{|l|}{} & JLF-S-2 & 79.39$\pm$0.17 & 81.43$\pm$0.16 & 76.69$\pm$0.25 & 77.16$\pm$0.30 & 79.98$\pm$0.25 & 75.89$\pm$0.11 & 76.54$\pm$1.25 & 79.97$\pm$1.23 & 74.73$\pm$1.16 \\ \hline
\multicolumn{1}{|l|}{\multirow{10}*{\rotatebox{90}{\textbf{Competitors}}}} & LF-MV & 78.41$\pm$0.32 & 80.09$\pm$0.31 & 76.24$\pm$0.22 & 75.48$\pm$0.29 & 77.98$\pm$0.20 & 73.82$\pm$0.21 & 74.88$\pm$1.13 & 77.64$\pm$1.14 & 73.15$\pm$1.11 \\ \cline{2-11}
\multicolumn{1}{|l|}{} & LF-P-1 & 77.38$\pm$0.24 & 79.87$\pm$0.22 & 75.18$\pm$0.37 & 74.65$\pm$0.19 & 76.67$\pm$0.30 & 72.96$\pm$0.15 & 73.85$\pm$1.15 & 76.56$\pm$1.22 & 72.06$\pm$1.35 \\ \cline{2-11}
\multicolumn{1}{|l|}{} & LF-P-2 & 77.37$\pm$0.29 & 79.78$\pm$0.28 & 75.11$\pm$0.35 & 74.60$\pm$0.16 & 76.58$\pm$0.34 & 72.88$\pm$0.09 & 73.74$\pm$1.15 & 76.53$\pm$1.24 & 72.00$\pm$1.31 \\ \cline{2-11}
\multicolumn{1}{|l|}{} & LF-D-1 & 77.40$\pm$0.20 & 79.95$\pm$0.27 & 75.22$\pm$0.30 & 74.67$\pm$0.25 & 76.60$\pm$0.31 & 72.94$\pm$0.03 & 73.80$\pm$1.25 & 76.55$\pm$1.26 & 72.09$\pm$1.22 \\ \cline{2-11}
\multicolumn{1}{|l|}{} & LF-D-2 & 77.34$\pm$0.18 & 79.90$\pm$0.29 & 75.20$\pm$0.29 & 74.58$\pm$0.24 & 76.49$\pm$0.27 & 72.87$\pm$0.24 & 73.71$\pm$1.22 & 76.39$\pm$1.23 & 72.00$\pm$1.26 \\ \cline{2-11}
\multicolumn{1}{|l|}{} & JF-C & 72.80$\pm$0.63 & 76.62$\pm$0.83 & 68.62$\pm$0.98 & 70.35$\pm$0.47 & 71.04$\pm$0.52 & 67.95$\pm$0.73 & 72.47$\pm$1.29 & 78.76$\pm$1.44 & 68.24$\pm$1.39 \\ \cline{2-11}
\multicolumn{1}{|l|}{} & JF-M & 75.00$\pm$0.76 & 78.30$\pm$0.71 & 71.47$\pm$0.85 & 71.37$\pm$0.49 & 74.59$\pm$0.65 & 70.47$\pm$0.49 & 71.64$\pm$1.37 & 75.06$\pm$1.41 & 69.90$\pm$1.83 \\ \cline{2-11}
\multicolumn{1}{|l|}{} & HC$^*$ & 75.50$\pm$0.70 & 75.80$\pm$0.80 & 75.30$\pm$1.30 & 75.20$\pm$6.70 & 71.10$\pm$16.50 & 82.40$\pm$15.40 & - & - & - \\ \cline{2-11}
\multicolumn{1}{|l|}{} & HYB$^*$ & 76.90$\pm$5.40 & 78.80$\pm$6.40 & 74.70$\pm$5.90 & 74.30$\pm$6.10 & 76.90$\pm$18.90 & 68.50$\pm$15.50 & - & - & - \\ \cline{2-11}
\multicolumn{1}{|l|}{} & ETE$^*$ & 74.80$\pm$0.80 & 74.50$\pm$1.70 & 75.10$\pm$1.50 & 70.90$\pm$0.50 & 73.40$\pm$1.80 & 69.60$\pm$0.90 & - & - & - \\ \hline
\end{tabular}}
\label{tab:joint}
\end{table}
The results reveal that the four joint-late methods, i.e. the first four rows in \tablename~\ref{tab:joint}, outperform the other techniques, and the performance differences against the other multimodal approaches are statistically significant in terms of accuracy, sensitivity and specificity.
In particular, these significant differences between JLF and LF-MV, LF-P and LF-D methods demonstrate that the end-to-end training of the method is effective and provide better results than the late fusion of the networks in the ensemble.
As mentioned in section~\ref{subsec:Competitors}, this result displays \textit{when} the models in the ensemble should be fused. 

Let us now focus on the comparison between the four JLF methods and the other two approaches providing an embedded multimodal representation, namely the JF-C, JF-M competitors described in section~\ref{sec:Background}. 
We find that the JLF performance are statically different in terms of accuracy, sensitivity and specificity with respect to those attained by the JF-C, JF-M techniques.
This finding points out \textit{how} the individual learners should be combined to provide a useful embedded multimodal representation.

We also found that the performance of JLF models are statistically different when compared with those presented in~\cite{bib:soda2021aiforcovid} (i.e. HC, HYB and ETE), which are the current baseline for the AIforCOVID dataset.

Turning our attention to the comparison between the four set-ups of the JLF approach, the results reveal that no statistically significant differences exist among their performance, suggesting that there is no preferred technique between them, while JFL-C-1 provides the best accuracy in both CV and LOCO experiments.

We are now interested in assessing to what extent our optimization algorithm effectively detects $\Gamma^*$ as the best choice of \textit{which} learners should be combined among all the possible $\Gamma_I$.
To this end, we randomly picked $100$ different $\Gamma_I$ that are joint using the JFL-C-1 technique for the reason mentioned a few lines above.
Next, we compare these performance against the four JLF results in~\tablename~\ref{tab:joint} that are achieved using $\Gamma^*$, and we find that they are statistically significantly lower.

We now focus on the unimodal scenario to investigate the performance in JLF when a modality is missing: to this end \tablename~\ref{tab:single} shows the results of single models in $\Gamma^*$ when trained on a single modality.
\begin{table}[]
\centering
\caption{Performance single modality models of $\Gamma^*$ in JLF}
\resizebox{\textwidth}{!}{
\begin{tabular}{|ll|ccc|ccc|ccc|}
\hline
\multirow{2}{*}{\textbf{Model}} & \multirow{2}{*}{\textbf{Data}} & \multicolumn{3}{c|}{\textbf{CV}} & \multicolumn{3}{c|}{\textbf{LOCO}} & \multicolumn{3}{c|}{\textbf{EV}} \\
 & & \textbf{Acc} & \textbf{TPR} & \textbf{TNR} & \textbf{Acc} & \textbf{TPR} & \textbf{TNR} & \textbf{Acc} & \textbf{TPR} & \textbf{TNR} \\ \hline
MLP-2 & CLI & 74.87$\pm$1.46 & 75.95$\pm$1.39 & 72.50$\pm$1.76 & 73.91$\pm$1.29 & 76.37$\pm$1.10 & 70.39$\pm$1.37 & 71.44$\pm$2.82 & 74.17$\pm$2.68 & 68.08$\pm$2.98 \\ \hline
GoogLeNet & IMG & 74.61$\pm$0.20 & 75.99$\pm$0.44 & 72.37$\pm$0.65 & 71.68$\pm$0.75 & 72.58$\pm$0.38 & 70.58$\pm$0.94 & 73.36$\pm$1.95 & 74.73$\pm$1.88 & 71.54$\pm$1.37 \\ \hline
VGG13-BN & IMG & 73.04$\pm$0.38 & 75.67$\pm$0.23 & 70.86$\pm$0.47 & 71.25$\pm$0.39 & 72.28$\pm$0.02 & 70.72$\pm$0.20 & 71.78$\pm$1.48 & 72.98$\pm$1.92 & 70.86$\pm$1.29 \\ \hline
ResNeXt50 & IMG & 72.91$\pm$0.86 & 73.90$\pm$0.87 & 71.07$\pm$0.38 & 67.43$\pm$0.28 & 70.08$\pm$0.10 & 65.60$\pm$0.10 & 73.33$\pm$1.65 & 75.73$\pm$1.10 & 70.62$\pm$1.28 \\ \hline
\end{tabular}}
\label{tab:single}
\end{table}
By performing the statistical assessments we find that there is no significant change with the unimodal results presented in~\cite{bib:soda2021aiforcovid}, as it could be expected.
Note also that for the sake of presentation in \tablename~\ref{tab:single} we do not report all the results attained by all the CNNs and MLPs considered.
Nevertheless, GoogLeNet, VGG13-BN and ResNeXt50 are actually the first, the second and the sixth, among the CNNs, respectively, whilst MLP-2 is the first among the other MLPs.
Comparing the four JLF learners with these unimodal networks, we find that the former provide performance that are always statistically larger than those returned by the latter, confirming the importance of the fusion between the modalities.

Let us now concentrate on a single modality.
Since $\Gamma^*$ is composed of three CNNs working on the image modality, we perform an ablation test removing the network using clinical data (i.e. MLP-2).
In this case, JLF-C-1 achieves an average accuracy equal to $76.09 \pm 0.29$, $75.68 \pm 0.31$ and $74.86 \pm 0.27$in CV, LOCO and EV, respectively.
Since these values are statistically lower than those of JLF, we deem that joining multiple networks is beneficial.
Furthermore, note that the apposite ablation test, i.e. removing the networks using image data, corresponds to consider MLP-2 only, whose performance are already shown in \tablename~\ref{tab:single}.

\begin{figure}[]
\centering
\caption{Distribution of the performance accuracy of all the presented models and competitors in all the training configuration scenarios.}
\includegraphics[width=\textwidth]{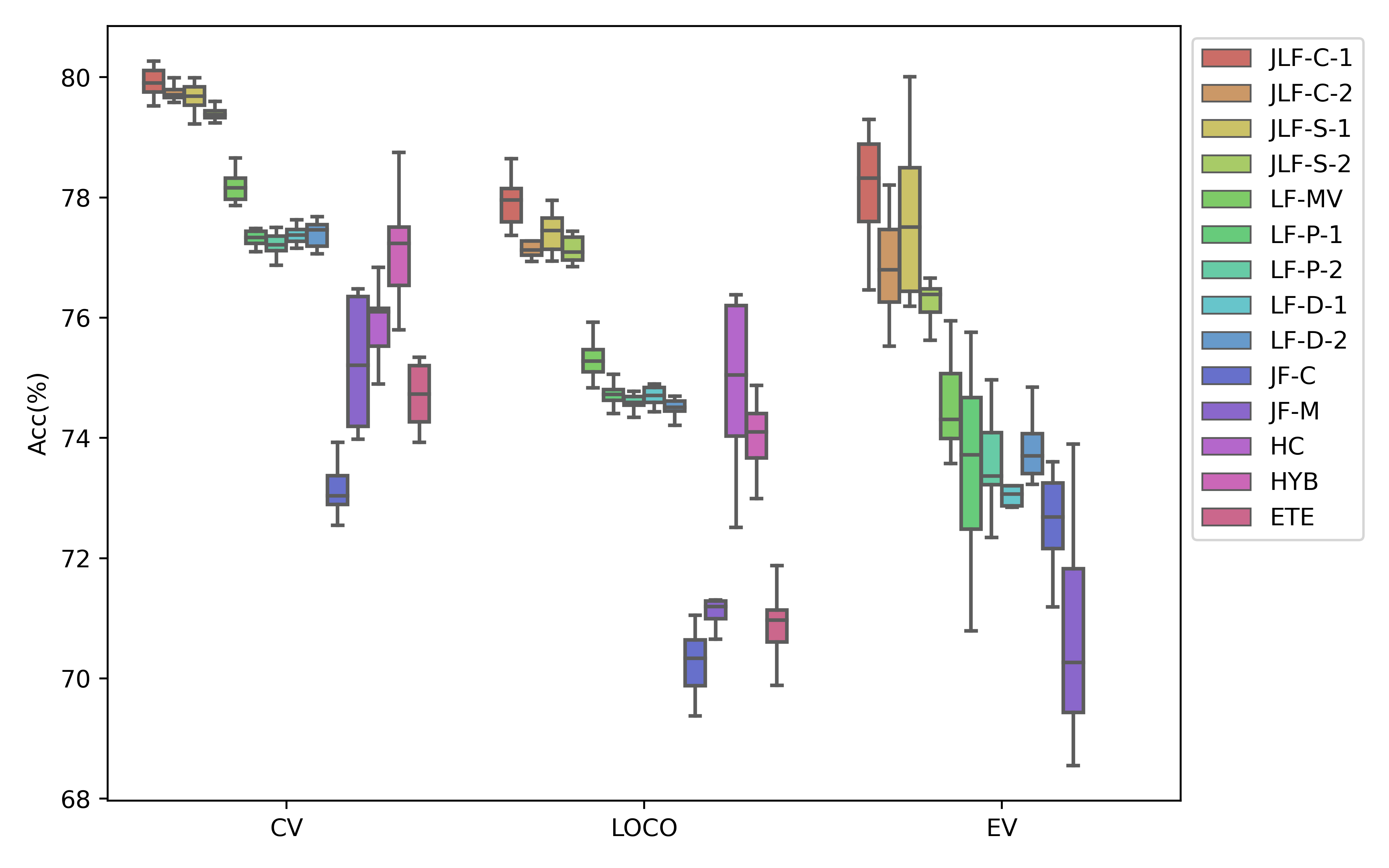}
\label{fig:results}
\end{figure}

As a further observation, let us now focus on the different performance achieved in 10-fold CV, LOCO and EV.
To make the comparison easier, \figurename~\ref{fig:results} shows the distribution of the performance of the different methods across all the experimental scenarios. 
Observing the results of the JLF methods, we notice that the method is robust and generalizable since the performance drop in LOCO and in EV is limited, while all the others suffer from a larger decrease.

\paragraph{Explainability}

To open the black-box nature of the multimodal deep approach presented in this work, we apply the XAI algorithms introduced in section~\ref{sec:Methods} to the model with the highest performance, i.e. JLF-C-1.

To this end, we first extract the weights coming out of the classification vector which are connected to the next hidden layer of the FC network; second, we compute the mean relative intensities of such weights attained in CV, obtaining the following distribution: 17\%, 26\%, 16\%, and 41\% for GoogLeNet, VGG13-BN, ResNeXt50, and MLP-2, respectively.
This information not only makes us understand the importance of every single model for the final classification, but it also explains the hierarchy inter- and intra-modality.
In particular, we notice that the image modality has more importance than the clinical one since the relative vector weights are 59\% and 41\%, respectively.
Even if there is a discrepancy between the number of models per modality, it is interesting to notice that MLP-2 has the highest importance among the single neural networks, but the combination of the three CNNs gives higher relative importance to the image modality.
Going deeper, we can see that there is also an intra-modality hierarchy between the image-based models, where by calculating the relative weight importances on GoogLeNet, VGG13-BN, and ResNeXt50, we obtain 29\%, 44\%, and 27\%, respectively.

\begin{figure}[]
\centering
\caption{XAI modality algorithms: a) feature importance of Integrated Gradients of an instance applied to MLP-2; b) activation maps of Grad-CAM on GoogLeNet, VGG13-BN, and ResNeXt50 which are combined resulting in the weighted activation map}
\includegraphics[width=\textwidth]{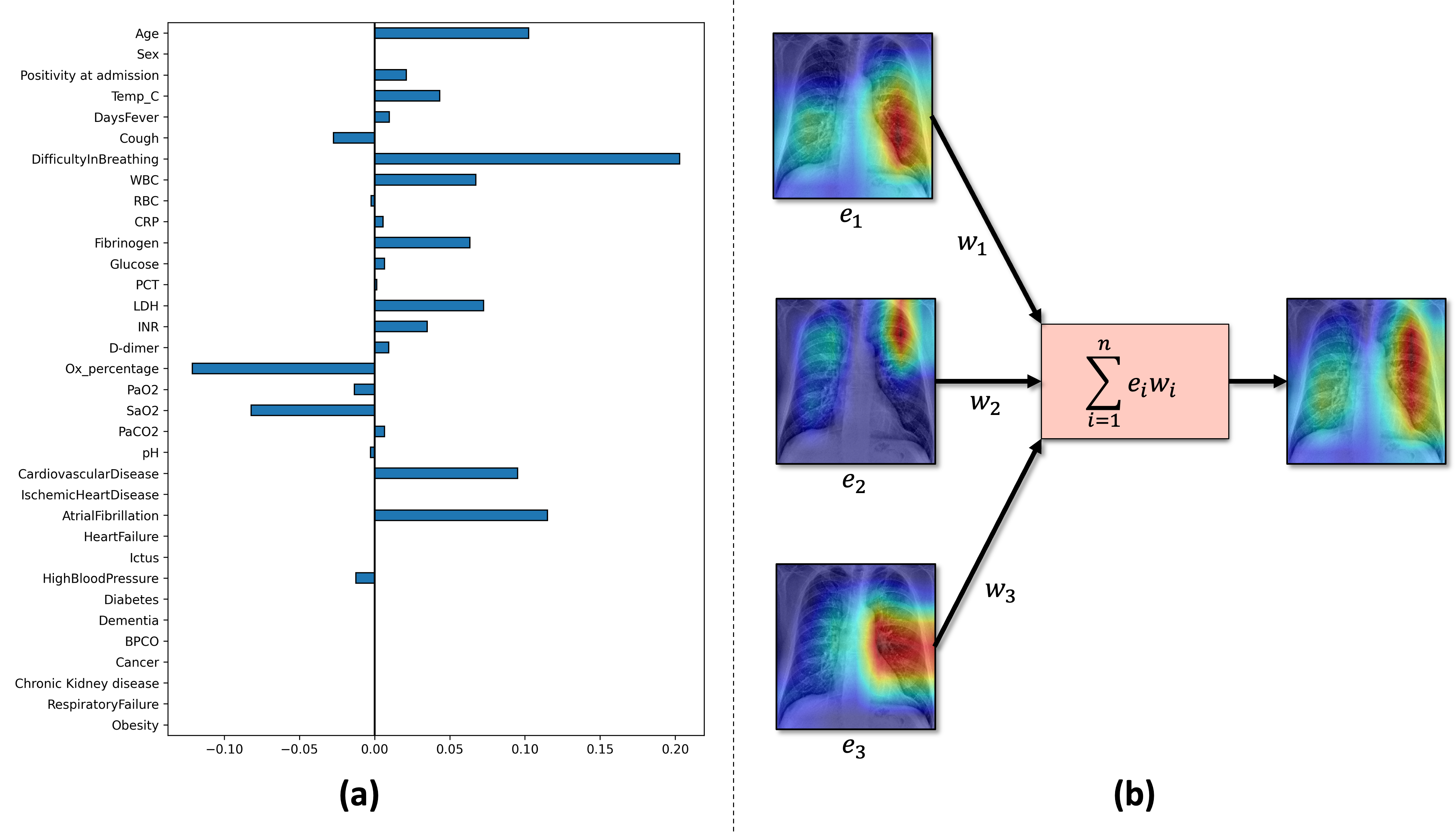}
\label{fig:XAI}
\end{figure}

To enable physicians to explore and understand data-driven DL-based systems, we work on XAI algorithms for single models. 
For each model composing the joint fusion, we apply XAI algorithms suited for the specific modality.
For example, \figurename~\ref{fig:XAI} shows the results of the integrated gradients algorithm~\cite{bib:sundararajan2017axiomatic} applied to MLP-2, which shows the importance of the clinical features for the classification.
We notice that the most important features are the difficulty in breathing and the oxygen percentage in blood.
Indeed, the presence of difficulty in breathing and lower values in the oxygen percentage in the blood are indications of the severe COVID-19 cases as also the medical literature confirms~\cite{bib:kermali2020role}.
Going forward, we can use the aforementioned relative weights of the classification vector for a specific modality to combine the results coming from an XAI algorithm.
Still in \figurename~\ref{fig:XAI}, we show the feature maps extracted from GoogLeNet, VGG13-BN, ResNeXt50 by applying Grad-CAM~\cite{bib:selvaraju2017grad}, and by combining them via equation~\ref{eq:weighted} we obtain a resulting map for the modality.
In this way we understand how the different components working on the image modality, and as a whole, interpret the pixel values.
Many more XAI algorithms can be applied following this methodology, but to prevent redundancy we limit ourselves to these examples.
We did not perform any evaluation on the XAI methods, since it is out of our scope of the presented method.

\section{Conclusions} \label{sec:Conclusions}

In this manuscript we have presented an approach to build an optimized multimodal end-to-end model exploiting a performance metric and the diversity of different neural networks. 
It addresses open issues in MDL related to \textit{when}, \textit{which} and \textit{how} the modalities and the related learners must be joined to provide effective embedded representation and satisfactory performance.
Given the impact of the COVID-19 pandemic, we applied our method to predict patients' prognosis, a topic that has attracted recent research after the large efforts towards the detection of COVID-19 signs in medical images~\cite{bib:wynants2020prediction}.
In this scenario, the use of different modalities capturing different aspects of the disease progression should be useful and, for this reason, we used the AIforCOVID dataset~\cite{bib:soda2021aiforcovid}, which is the only publicly available dataset containing clinical data and CXR scans.
Our proposal algorithmically determines \textit{which} deep architectures for each modality should be fused.
Furthermore, here we have investigated \textit{how} and \textit{when} the fusion of the modalities should occur: we find that concatenating the classification vectors of each learner, being either soft or crisp labels, is a viable solution.
In terms of performance, our method attains state-of-the-art results, not only outperforming the baseline performance reported in~\cite{bib:soda2021aiforcovid} but also being robust to external validation.
Moreover, our method fits well with many XAI algorithms, which allow us to figure out a hierarchy among modalities and they can extract the intra-modality importance of the different features by combining the results of various networks.

Future directions are directed towards three main goals.
First, we plan to study a method to minimize the computational costs to extend the number of possible architectures per modality in the case of end-to-end training.
Second, to investigate the application of this approach to other multimodal datasets.
Third, as medical data should come from different institutions, it would also be interesting to explore the deployment of our approach in a federated learning framework.

\section*{Acknowledgments}
This work is partially funded by: POR CAMPANIA FESR 2014 - 2020, AP1-OS1.3 project ``Protocolli TC del torace a bassissima dose e tecniche di intelligenza artificiale per la diagnosi precoce e quantificazione della malattia da COVID-19'' CUP D54I20001410002; EU project ``University-Industrial Educational Centre in Advanced Biomedical and Medical Informatics (CeBMI) No. 612462-EPP-1-2019-1-SK-EPPKA2-KA''.
The authors would like to thank the team collecting and making publicly available the AIforCOVID dataset.

\appendix
\setcounter{table}{0}
\section{Clinical data}

\begin{table}[h]
\centering
\caption{Description of the clinical data coming from~\cite{bib:soda2021aiforcovid}. Feature names followed by $+$ were not used.}
\setlength\extrarowheight{-0.25pt}
\resizebox{\textwidth}{!}{
\begin{tabular}{|ll|}
\hline
\textbf{Name} & \textbf{Description} \\ \hline
Active cancer in the last 5 years & Patient had active cancer in the last 5 years\\ \hline
Age & Patient's age (years) \\\hline
Atrial Fibrillation & Patient had atrial fibrillaton\\ \hline
Body temperature ($^{\circ}$C) & Patients temperature at admission (in $^{\circ}$C) \\ \hline
Cardiovascular Disease & Patient had cardiovascular diseases\\ \hline
Chronic Kidney disease & Patient had chronic kidney disease\\ \hline
COPD & Chronic obstructive pulmonary disease \\ \hline
Cough & Cough presence \\ \hline
CRP & C-reactive protein concentration (mg/dL) \\ \hline
Days Fever & Days of fever up to admission (days) \\ \hline
D-dimer & D-dimer amount in blood \\ \hline
Death+ & Death of patient occurred during hospitalization for any cause \\ \hline
Dementia & Patient had dementia\\ \hline
Diabetes & Patient had diabetes\\ \hline
Dyspnea & Patient had intense tightening in the chest, air hunger, difficulty breathing, breathlessness or a feeling of suffocation\\ \hline
Fibrinogen & Fibrinogen concentration in blood (mg/dL) \\ \hline
Glucose & Glucose concentration in blood (mg/dL) \\ \hline
Heart Failure & Patient had heart failure\\ \hline
Hypertension & Patient had high blood pressure \\ \hline
INR & International Normalized Ratio \\\hline 
Ischemic Heart Disease & Patient had ischemic heart disease \\ \hline
LDH & Lactate dehydrogenase concentration in blood (U/L) \\ \hline
$O_2$ (\%) & Oxygen percentage in blood (\%) \\ \hline
Obesity & Patient had obesity \\ \hline
$PaCO_2$ & Partial pressure of carbon dioxide in arterial blood (mmHg) \\ \hline
$PaO_2$ & Partial pressure of oxygen in arterial blood (mmHg) \\ \hline
PCT & Platelet count (ng/mL) \\ \hline
pH & Blood pH \\ \hline
Position+ & Patient position during chest x-ray (\%supine) \\\hline
Positivity at admission & Positivity to the SARS-CoV-2 swab at the admission time\\ \hline
Prognosis & Patient outcome \\ \hline
RBC & Red blood cells count (10\textasciicircum{}9/L) \\ \hline
Respiratory Failure & Patient had respiratory failure \\ \hline
$SaO_2$ & Arterial oxygen saturation (\%) \\ \hline
Sex & Patient's sex \\ \hline
Stroke & Patient had stroke \\ \hline
Therapy Anakinra+ & Patient was treated with Anakinra \\ \hline
Therapy anti-inflammatory+ & Patient was treated with anti-inflammatory drugs therapy \\ \hline
Therapy antiviral$+$ & Patient was treated with antiviral drugs \\ \hline
Therapy Eparine$+$ & Patient was treated with eparine treatment; therapeutic treatment \\ \hline
Therapy hydroxychloroquine$+$ & Patient was treated with hydroxychloroquine \\\hline
Therapy Tocilizumab$+$ & Patient was treated with Tocilizumab \\ \hline
WBC & White blood cells count (10\textasciicircum{}9/L) \\ \hline
\end{tabular}}
\label{tab:clinicaldata}
\end{table}


\begin{thebibliography}{55}
\expandafter\ifx\csname natexlab\endcsname\relax\def\natexlab#1{#1}\fi
\providecommand{\url}[1]{\texttt{#1}}
\providecommand{\href}[2]{#2}
\providecommand{\path}[1]{#1}
\providecommand{\DOIprefix}{doi:}
\providecommand{\ArXivprefix}{arXiv:}
\providecommand{\URLprefix}{URL: }
\providecommand{\Pubmedprefix}{pmid:}
\providecommand{\doi}[1]{\href{http://dx.doi.org/#1}{\path{#1}}}
\providecommand{\Pubmed}[1]{\href{pmid:#1}{\path{#1}}}
\providecommand{\bibinfo}[2]{#2}
\ifx\xfnm\relax \def\xfnm[#1]{\unskip,\space#1}\fi
\bibitem[{Al-Najjar and Al-Rousan(2020)}]{bib:al2020classifier}
\bibinfo{author}{Al-Najjar, H.}, \bibinfo{author}{Al-Rousan, N.},
  \bibinfo{year}{2020}.
\newblock \bibinfo{title}{{A classifier prediction model to predict the status
  of Coronavirus {COVID-19} patients in South Korea}}.
\newblock \bibinfo{journal}{European Review for Medical and Pharmacological
  Sciences} .
\bibitem[{Arcuri and Fraser(2013)}]{arcuri2013parameter}
\bibinfo{author}{Arcuri, A.}, \bibinfo{author}{Fraser, G.},
  \bibinfo{year}{2013}.
\newblock \bibinfo{title}{{Parameter tuning or default values? An empirical
  investigation in search-based software engineering}}.
\newblock \bibinfo{journal}{Empirical Software Engineering}
  \bibinfo{volume}{18}, \bibinfo{pages}{594--623}.
\bibitem[{Arrieta et~al.(2020)Arrieta, D{\'\i}az-Rodr{\'\i}guez, Del~Ser,
  Bennetot, Tabik, Barbado, Garc{\'\i}a, Gil-L{\'o}pez, Molina, Benjamins
  et~al.}]{bib:arrieta2020explainable}
\bibinfo{author}{Arrieta, A.B.}, \bibinfo{author}{D{\'\i}az-Rodr{\'\i}guez,
  N.}, \bibinfo{author}{Del~Ser, J.}, \bibinfo{author}{Bennetot, A.},
  \bibinfo{author}{Tabik, S.}, \bibinfo{author}{Barbado, A.},
  \bibinfo{author}{Garc{\'\i}a, S.}, \bibinfo{author}{Gil-L{\'o}pez, S.},
  \bibinfo{author}{Molina, D.}, \bibinfo{author}{Benjamins, R.}, et~al.,
  \bibinfo{year}{2020}.
\newblock \bibinfo{title}{Explainable artificial intelligence {(XAI)}:
  Concepts, taxonomies, opportunities and challenges toward responsible {AI}}.
\newblock \bibinfo{journal}{Information Fusion} \bibinfo{volume}{58},
  \bibinfo{pages}{82--115}.
\bibitem[{Bai et~al.(2020)Bai, Fang, Zhou, Bai, Liu, Xia, Chen, Xu, Xia, Gong
  et~al.}]{bib:bai2020predicting}
\bibinfo{author}{Bai, X.}, \bibinfo{author}{Fang, C.}, \bibinfo{author}{Zhou,
  Y.}, \bibinfo{author}{Bai, S.}, \bibinfo{author}{Liu, Z.},
  \bibinfo{author}{Xia, L.}, \bibinfo{author}{Chen, Q.}, \bibinfo{author}{Xu,
  Y.}, \bibinfo{author}{Xia, T.}, \bibinfo{author}{Gong, S.}, et~al.,
  \bibinfo{year}{2020}.
\newblock \bibinfo{title}{Predicting {COVID-19} malignant progression with {AI}
  techniques} .
\bibitem[{Baltru{\v{s}}aitis et~al.(2018)Baltru{\v{s}}aitis, Ahuja and
  Morency}]{bib:baltruvsaitis2018multimodal}
\bibinfo{author}{Baltru{\v{s}}aitis, T.}, \bibinfo{author}{Ahuja, C.},
  \bibinfo{author}{Morency, L.P.}, \bibinfo{year}{2018}.
\newblock \bibinfo{title}{Multimodal machine learning: A survey and taxonomy}.
\newblock \bibinfo{journal}{IEEE transactions on pattern analysis and machine
  intelligence} \bibinfo{volume}{41}, \bibinfo{pages}{423--443}.
\bibitem[{Bengio et~al.(2013)Bengio, Courville and
  Vincent}]{bib:bengio2013representation}
\bibinfo{author}{Bengio, Y.}, \bibinfo{author}{Courville, A.},
  \bibinfo{author}{Vincent, P.}, \bibinfo{year}{2013}.
\newblock \bibinfo{title}{Representation learning: A review and new
  perspectives}.
\newblock \bibinfo{journal}{IEEE Transactions on Pattern Analysis and Machine
  Intelligence} \bibinfo{volume}{35}, \bibinfo{pages}{1798--1828}.
\bibitem[{Chamola et~al.(2020)Chamola, Hassija, Gupta and
  Guizani}]{bib:chamola2020comprehensive}
\bibinfo{author}{Chamola, V.}, \bibinfo{author}{Hassija, V.},
  \bibinfo{author}{Gupta, V.}, \bibinfo{author}{Guizani, M.},
  \bibinfo{year}{2020}.
\newblock \bibinfo{title}{A comprehensive review of the {COVID-19} pandemic and
  the role of {IoT}, drones, {AI}, blockchain, and {5G} in managing its
  impact}.
\newblock \bibinfo{journal}{IEEE Access} \bibinfo{volume}{8},
  \bibinfo{pages}{90225--90265}.
\bibitem[{Chen et~al.(2020)Chen, Lu, Wang, Williamson, Rodig, Lindeman and
  Mahmood}]{bib:chen2020pathomic}
\bibinfo{author}{Chen, R.J.}, \bibinfo{author}{Lu, M.Y.},
  \bibinfo{author}{Wang, J.}, \bibinfo{author}{Williamson, D.F.},
  \bibinfo{author}{Rodig, S.J.}, \bibinfo{author}{Lindeman, N.I.},
  \bibinfo{author}{Mahmood, F.}, \bibinfo{year}{2020}.
\newblock \bibinfo{title}{Pathomic fusion: an integrated framework for fusing
  histopathology and genomic features for cancer diagnosis and prognosis}.
\newblock \bibinfo{journal}{IEEE Transactions on Medical Imaging} .
\bibitem[{Cohen et~al.(2020)Cohen, Morrison, Dao, Roth, Duong and
  Ghassemi}]{bib:cohen2020covid}
\bibinfo{author}{Cohen, J.P.}, \bibinfo{author}{Morrison, P.},
  \bibinfo{author}{Dao, L.}, \bibinfo{author}{Roth, K.},
  \bibinfo{author}{Duong, T.Q.}, \bibinfo{author}{Ghassemi, M.},
  \bibinfo{year}{2020}.
\newblock \bibinfo{title}{{COVID-19} image data collection: Prospective
  predictions are the future}.
\newblock \bibinfo{journal}{arXiv preprint arXiv:2006.11988} .
\bibitem[{Deng et~al.(2009)Deng, Dong, Socher, Li, Li and
  Fei-Fei}]{bib:deng2009imagenet}
\bibinfo{author}{Deng, J.}, \bibinfo{author}{Dong, W.},
  \bibinfo{author}{Socher, R.}, \bibinfo{author}{Li, L.J.},
  \bibinfo{author}{Li, K.}, \bibinfo{author}{Fei-Fei, L.},
  \bibinfo{year}{2009}.
\newblock \bibinfo{title}{{ImageNet}: A large-scale hierarchical image
  database}, in: \bibinfo{booktitle}{2009 {IEEE} Conference on Computer Vision
  and Pattern Recognition}, \bibinfo{organization}{{IEEE}}. pp.
  \bibinfo{pages}{248--255}.
\bibitem[{Fang et~al.(2021)Fang, Bai, Chen, Zhou, Xia, Qin, Gong, Xie, Zhou, Tu
  et~al.}]{bib:fang2021deep}
\bibinfo{author}{Fang, C.}, \bibinfo{author}{Bai, S.}, \bibinfo{author}{Chen,
  Q.}, \bibinfo{author}{Zhou, Y.}, \bibinfo{author}{Xia, L.},
  \bibinfo{author}{Qin, L.}, \bibinfo{author}{Gong, S.}, \bibinfo{author}{Xie,
  X.}, \bibinfo{author}{Zhou, C.}, \bibinfo{author}{Tu, D.}, et~al.,
  \bibinfo{year}{2021}.
\newblock \bibinfo{title}{Deep learning for predicting covid-19 malignant
  progression}.
\newblock \bibinfo{journal}{Medical Image Analysis} \bibinfo{volume}{72},
  \bibinfo{pages}{102096}.
\bibitem[{Glorot and Bengio(2010)}]{bib:glorot2010understanding}
\bibinfo{author}{Glorot, X.}, \bibinfo{author}{Bengio, Y.},
  \bibinfo{year}{2010}.
\newblock \bibinfo{title}{Understanding the difficulty of training deep
  feedforward neural networks}, in: \bibinfo{booktitle}{Proceedings of the
  Thirteenth International Conference on Artificial Intelligence and
  Statistics}, \bibinfo{organization}{JMLR Workshop and Conference
  Proceedings}. pp. \bibinfo{pages}{249--256}.
\bibitem[{Glorot et~al.(2011)Glorot, Bordes and Bengio}]{bib:glorot2011deep}
\bibinfo{author}{Glorot, X.}, \bibinfo{author}{Bordes, A.},
  \bibinfo{author}{Bengio, Y.}, \bibinfo{year}{2011}.
\newblock \bibinfo{title}{Deep sparse rectifier neural networks}, in:
  \bibinfo{booktitle}{Proceedings of the Fourteenth International Conference on
  Artificial Intelligence and Statistics}, \bibinfo{organization}{JMLR Workshop
  and Conference Proceedings}. pp. \bibinfo{pages}{315--323}.
\bibitem[{Guarrasi et~al.(2021)Guarrasi, D'Amico, Sicilia, Cordelli and
  Soda}]{bib:guarrasi2021multi}
\bibinfo{author}{Guarrasi, V.}, \bibinfo{author}{D'Amico, N.C.},
  \bibinfo{author}{Sicilia, R.}, \bibinfo{author}{Cordelli, E.},
  \bibinfo{author}{Soda, P.}, \bibinfo{year}{2021}.
\newblock \bibinfo{title}{A multi-expert system to detect covid-19 cases in
  x-ray images}, in: \bibinfo{booktitle}{2021 IEEE 34th International Symposium
  on Computer-Based Medical Systems (CBMS)}, \bibinfo{organization}{IEEE}. pp.
  \bibinfo{pages}{395--400}.
\bibitem[{Guarrasi et~al.(2022)Guarrasi, D’Amico, Sicilia, Cordelli and
  Soda}]{bib:guarrasi2022pareto}
\bibinfo{author}{Guarrasi, V.}, \bibinfo{author}{D’Amico, N.C.},
  \bibinfo{author}{Sicilia, R.}, \bibinfo{author}{Cordelli, E.},
  \bibinfo{author}{Soda, P.}, \bibinfo{year}{2022}.
\newblock \bibinfo{title}{Pareto optimization of deep networks for {COVID-19}
  diagnosis from chest {X-rays}}.
\newblock \bibinfo{journal}{Pattern Recognition} \bibinfo{volume}{121},
  \bibinfo{pages}{108242}.
\bibitem[{He et~al.(2016)He, Zhang, Ren and Sun}]{bib:he2016deep}
\bibinfo{author}{He, K.}, \bibinfo{author}{Zhang, X.}, \bibinfo{author}{Ren,
  S.}, \bibinfo{author}{Sun, J.}, \bibinfo{year}{2016}.
\newblock \bibinfo{title}{Deep residual learning for image recognition}, in:
  \bibinfo{booktitle}{Proceedings of the IEEE Conference on Computer Vision and
  Pattern Recognition}, pp. \bibinfo{pages}{770--778}.
\bibitem[{Hendricks et~al.(2018)Hendricks, Burns, Saenko, Darrell and
  Rohrbach}]{bib:hendricks2018women}
\bibinfo{author}{Hendricks, L.A.}, \bibinfo{author}{Burns, K.},
  \bibinfo{author}{Saenko, K.}, \bibinfo{author}{Darrell, T.},
  \bibinfo{author}{Rohrbach, A.}, \bibinfo{year}{2018}.
\newblock \bibinfo{title}{Women also snowboard: Overcoming bias in captioning
  models}, in: \bibinfo{booktitle}{Proceedings of the European Conference on
  Computer Vision (ECCV)}, pp. \bibinfo{pages}{771--787}.
\bibitem[{Huang et~al.(2017)Huang, Liu, Van Der~Maaten and
  Weinberger}]{bib:huang2017densely}
\bibinfo{author}{Huang, G.}, \bibinfo{author}{Liu, Z.}, \bibinfo{author}{Van
  Der~Maaten, L.}, \bibinfo{author}{Weinberger, K.Q.}, \bibinfo{year}{2017}.
\newblock \bibinfo{title}{Densely connected convolutional networks}, in:
  \bibinfo{booktitle}{Proceedings of the IEEE Conference on Computer Vision and
  Pattern Recognition}, pp. \bibinfo{pages}{4700--4708}.
\bibitem[{Jaeger et~al.(2014)Jaeger, Candemir, Antani, W{\'a}ng, Lu and
  Thoma}]{bib:jaeger2014two}
\bibinfo{author}{Jaeger, S.}, \bibinfo{author}{Candemir, S.},
  \bibinfo{author}{Antani, S.}, \bibinfo{author}{W{\'a}ng, Y.X.J.},
  \bibinfo{author}{Lu, P.X.}, \bibinfo{author}{Thoma, G.},
  \bibinfo{year}{2014}.
\newblock \bibinfo{title}{Two public chest {X-ray} datasets for computer-aided
  screening of pulmonary diseases}.
\newblock \bibinfo{journal}{Quantitative Imaging in Medicine and Surgery}
  \bibinfo{volume}{4}, \bibinfo{pages}{475}.
\bibitem[{John(2014)}]{bib:john2014extremum}
\bibinfo{author}{John, F.}, \bibinfo{year}{2014}.
\newblock \bibinfo{title}{Extremum problems with inequalities as subsidiary
  conditions}, in: \bibinfo{booktitle}{Traces and Emergence of Nonlinear
  Programming}. \bibinfo{publisher}{Springer}, pp. \bibinfo{pages}{197--215}.
\bibitem[{Joshi et~al.(2021)Joshi, Walambe and Kotecha}]{bib:joshi2021review}
\bibinfo{author}{Joshi, G.}, \bibinfo{author}{Walambe, R.},
  \bibinfo{author}{Kotecha, K.}, \bibinfo{year}{2021}.
\newblock \bibinfo{title}{A review on explainability in multimodal deep neural
  nets}.
\newblock \bibinfo{journal}{IEEE Access} .
\bibitem[{Joze et~al.(2020)Joze, Shaban, Iuzzolino and
  Koishida}]{bib:joze2020mmtm}
\bibinfo{author}{Joze, H.R.V.}, \bibinfo{author}{Shaban, A.},
  \bibinfo{author}{Iuzzolino, M.L.}, \bibinfo{author}{Koishida, K.},
  \bibinfo{year}{2020}.
\newblock \bibinfo{title}{{MMTM}: Multimodal transfer module for {CNN} fusion},
  in: \bibinfo{booktitle}{Proceedings of the IEEE/CVF Conference on Computer
  Vision and Pattern Recognition}, pp. \bibinfo{pages}{13289--13299}.
\bibitem[{Karpathy et~al.(2014)Karpathy, Toderici, Shetty, Leung, Sukthankar
  and Fei-Fei}]{bib:karpathy2014large}
\bibinfo{author}{Karpathy, A.}, \bibinfo{author}{Toderici, G.},
  \bibinfo{author}{Shetty, S.}, \bibinfo{author}{Leung, T.},
  \bibinfo{author}{Sukthankar, R.}, \bibinfo{author}{Fei-Fei, L.},
  \bibinfo{year}{2014}.
\newblock \bibinfo{title}{Large-scale video classification with convolutional
  neural networks}, in: \bibinfo{booktitle}{Proceedings of the IEEE Conference
  on Computer Vision and Pattern Recognition}, pp. \bibinfo{pages}{1725--1732}.
\bibitem[{Kermali et~al.(2020)Kermali, Khalsa, Pillai, Ismail and
  Harky}]{bib:kermali2020role}
\bibinfo{author}{Kermali, M.}, \bibinfo{author}{Khalsa, R.K.},
  \bibinfo{author}{Pillai, K.}, \bibinfo{author}{Ismail, Z.},
  \bibinfo{author}{Harky, A.}, \bibinfo{year}{2020}.
\newblock \bibinfo{title}{The role of biomarkers in diagnosis of {COVID-19}: A
  systematic review}.
\newblock \bibinfo{journal}{Life Sciences} \bibinfo{volume}{254},
  \bibinfo{pages}{117788}.
\bibitem[{Kim and Frahm(2018)}]{bib:kim2018hierarchy}
\bibinfo{author}{Kim, H.J.}, \bibinfo{author}{Frahm, J.M.},
  \bibinfo{year}{2018}.
\newblock \bibinfo{title}{Hierarchy of alternating specialists for scene
  recognition}, in: \bibinfo{booktitle}{Proceedings of the European Conference
  on Computer Vision (ECCV)}, pp. \bibinfo{pages}{451--467}.
\bibitem[{Krizhevsky(2014)}]{bib:krizhevsky2014one}
\bibinfo{author}{Krizhevsky, A.}, \bibinfo{year}{2014}.
\newblock \bibinfo{title}{One weird trick for parallelizing convolutional
  neural networks}.
\newblock \bibinfo{journal}{{arXiv} preprint {arXiv}:1404.5997} .
\bibitem[{Kuhn and Tucker(2014)}]{bib:kuhn2014nonlinear}
\bibinfo{author}{Kuhn, H.W.}, \bibinfo{author}{Tucker, A.W.},
  \bibinfo{year}{2014}.
\newblock \bibinfo{title}{Nonlinear programming}, in:
  \bibinfo{booktitle}{Traces and Emergence of Nonlinear Programming}.
  \bibinfo{publisher}{Springer}, pp. \bibinfo{pages}{247--258}.
\bibitem[{Kumar and Minz(2014)}]{bib:kumar2014feature}
\bibinfo{author}{Kumar, V.}, \bibinfo{author}{Minz, S.}, \bibinfo{year}{2014}.
\newblock \bibinfo{title}{Feature selection: a literature review}.
\newblock \bibinfo{journal}{SmartCR} \bibinfo{volume}{4},
  \bibinfo{pages}{211--229}.
\bibitem[{Kuncheva and Whitaker(2003)}]{bib:kuncheva2003measures}
\bibinfo{author}{Kuncheva, L.I.}, \bibinfo{author}{Whitaker, C.J.},
  \bibinfo{year}{2003}.
\newblock \bibinfo{title}{Measures of diversity in classifier ensembles and
  their relationship with the ensemble accuracy}.
\newblock \bibinfo{journal}{Machine Learning} \bibinfo{volume}{51},
  \bibinfo{pages}{181--207}.
\bibitem[{Ma et~al.(2018)Ma, Zhang, Zheng and Sun}]{bib:ma2018shufflenet}
\bibinfo{author}{Ma, N.}, \bibinfo{author}{Zhang, X.}, \bibinfo{author}{Zheng,
  H.T.}, \bibinfo{author}{Sun, J.}, \bibinfo{year}{2018}.
\newblock \bibinfo{title}{{ShuffleNet V2}: Practical guidelines for efficient
  {CNN} architecture design}, in: \bibinfo{booktitle}{Proceedings of the
  European Conference on Computer Vision (ECCV)}, pp.
  \bibinfo{pages}{116--131}.
\bibitem[{Montavon et~al.(2017)Montavon, Lapuschkin, Binder, Samek and
  M{\"u}ller}]{bib:montavon2017explaining}
\bibinfo{author}{Montavon, G.}, \bibinfo{author}{Lapuschkin, S.},
  \bibinfo{author}{Binder, A.}, \bibinfo{author}{Samek, W.},
  \bibinfo{author}{M{\"u}ller, K.R.}, \bibinfo{year}{2017}.
\newblock \bibinfo{title}{Explaining nonlinear classification decisions with
  deep taylor decomposition}.
\newblock \bibinfo{journal}{Pattern Recognition} \bibinfo{volume}{65},
  \bibinfo{pages}{211--222}.
\bibitem[{Neverova et~al.(2015)Neverova, Wolf, Taylor and
  Nebout}]{bib:neverova2015moddrop}
\bibinfo{author}{Neverova, N.}, \bibinfo{author}{Wolf, C.},
  \bibinfo{author}{Taylor, G.}, \bibinfo{author}{Nebout, F.},
  \bibinfo{year}{2015}.
\newblock \bibinfo{title}{{ModDrop}: adaptive multi-modal gesture recognition}.
\newblock \bibinfo{journal}{IEEE Transactions on Pattern Analysis and Machine
  Intelligence} \bibinfo{volume}{38}, \bibinfo{pages}{1692--1706}.
\bibitem[{Ngiam et~al.(2011)Ngiam, Khosla, Kim, Nam, Lee and
  Ng}]{bib:ngiam2011multimodal}
\bibinfo{author}{Ngiam, J.}, \bibinfo{author}{Khosla, A.},
  \bibinfo{author}{Kim, M.}, \bibinfo{author}{Nam, J.}, \bibinfo{author}{Lee,
  H.}, \bibinfo{author}{Ng, A.Y.}, \bibinfo{year}{2011}.
\newblock \bibinfo{title}{Multimodal deep learning}, in:
  \bibinfo{booktitle}{International Conference on Machine Learning}.
\bibitem[{Ning et~al.(2020)Ning, Lei, Yang, Cao, Jiang, Yang, Zhang, Wang,
  Chen, Geng et~al.}]{bib:ning2020open}
\bibinfo{author}{Ning, W.}, \bibinfo{author}{Lei, S.}, \bibinfo{author}{Yang,
  J.}, \bibinfo{author}{Cao, Y.}, \bibinfo{author}{Jiang, P.},
  \bibinfo{author}{Yang, Q.}, \bibinfo{author}{Zhang, J.},
  \bibinfo{author}{Wang, X.}, \bibinfo{author}{Chen, F.},
  \bibinfo{author}{Geng, Z.}, et~al., \bibinfo{year}{2020}.
\newblock \bibinfo{title}{Open resource of clinical data from patients with
  pneumonia for the prediction of {COVID-19} outcomes via deep learning}.
\newblock \bibinfo{journal}{Nature Biomedical Engineering} \bibinfo{volume}{4},
  \bibinfo{pages}{1197--1207}.
\bibitem[{Nojavanasghari et~al.(2016)Nojavanasghari, Gopinath, Koushik,
  Baltru{\v{s}}aitis and Morency}]{bib:nojavanasghari2016deep}
\bibinfo{author}{Nojavanasghari, B.}, \bibinfo{author}{Gopinath, D.},
  \bibinfo{author}{Koushik, J.}, \bibinfo{author}{Baltru{\v{s}}aitis, T.},
  \bibinfo{author}{Morency, L.P.}, \bibinfo{year}{2016}.
\newblock \bibinfo{title}{Deep multimodal fusion for persuasiveness
  prediction}, in: \bibinfo{booktitle}{Proceedings of the 18th ACM
  International Conference on Multimodal Interaction}, pp.
  \bibinfo{pages}{284--288}.
\bibitem[{Ozsahin et~al.(2020)Ozsahin, Sekeroglu, Musa, Mustapha and
  Uzun~Ozsahin}]{bib:ozsahin2020review}
\bibinfo{author}{Ozsahin, I.}, \bibinfo{author}{Sekeroglu, B.},
  \bibinfo{author}{Musa, M.S.}, \bibinfo{author}{Mustapha, M.T.},
  \bibinfo{author}{Uzun~Ozsahin, D.}, \bibinfo{year}{2020}.
\newblock \bibinfo{title}{Review on diagnosis of {COVID-19} from chest {CT}
  images using artificial intelligence}.
\newblock \bibinfo{journal}{Computational and Mathematical Methods in Medicine}
  \bibinfo{volume}{2020}.
\bibitem[{Rajaraman et~al.(2020)Rajaraman, Siegelman, Alderson, Folio, Folio
  and Antani}]{rajaraman2020iteratively}
\bibinfo{author}{Rajaraman, S.}, \bibinfo{author}{Siegelman, J.},
  \bibinfo{author}{Alderson, P.O.}, \bibinfo{author}{Folio, L.S.},
  \bibinfo{author}{Folio, L.R.}, \bibinfo{author}{Antani, S.K.},
  \bibinfo{year}{2020}.
\newblock \bibinfo{title}{{Iteratively pruned deep learning ensembles for
  COVID-19 detection in chest X-rays}}.
\newblock \bibinfo{journal}{IEEE Access} \bibinfo{volume}{8},
  \bibinfo{pages}{115041--115050}.
\bibitem[{Ramachandram et~al.(2018)Ramachandram, Lisicki, Shields, Amer and
  Taylor}]{bib:ramachandram2018bayesian}
\bibinfo{author}{Ramachandram, D.}, \bibinfo{author}{Lisicki, M.},
  \bibinfo{author}{Shields, T.J.}, \bibinfo{author}{Amer, M.R.},
  \bibinfo{author}{Taylor, G.W.}, \bibinfo{year}{2018}.
\newblock \bibinfo{title}{Bayesian optimization on graph-structured search
  spaces: Optimizing deep multimodal fusion architectures}.
\newblock \bibinfo{journal}{Neurocomputing} \bibinfo{volume}{298},
  \bibinfo{pages}{80--89}.
\bibitem[{Ramachandram and Taylor(2017)}]{bib:ramachandram2017deep}
\bibinfo{author}{Ramachandram, D.}, \bibinfo{author}{Taylor, G.W.},
  \bibinfo{year}{2017}.
\newblock \bibinfo{title}{Deep multimodal learning: A survey on recent advances
  and trends}.
\newblock \bibinfo{journal}{IEEE Signal Processing Magazine}
  \bibinfo{volume}{34}, \bibinfo{pages}{96--108}.
\bibitem[{Ronneberger et~al.(2015)Ronneberger, Fischer and
  Brox}]{bib:ronneberger2015u}
\bibinfo{author}{Ronneberger, O.}, \bibinfo{author}{Fischer, P.},
  \bibinfo{author}{Brox, T.}, \bibinfo{year}{2015}.
\newblock \bibinfo{title}{{U-Net}: Convolutional networks for biomedical image
  segmentation}, in: \bibinfo{booktitle}{International Conference on Medical
  Image Computing and Computer-Assisted Intervention},
  \bibinfo{organization}{Springer}. pp. \bibinfo{pages}{234--241}.
\bibitem[{Sandler et~al.(2018)Sandler, Howard, Zhu, Zhmoginov and
  Chen}]{bib:sandler2018mobilenetv2}
\bibinfo{author}{Sandler, M.}, \bibinfo{author}{Howard, A.},
  \bibinfo{author}{Zhu, M.}, \bibinfo{author}{Zhmoginov, A.},
  \bibinfo{author}{Chen, L.C.}, \bibinfo{year}{2018}.
\newblock \bibinfo{title}{{MobileNetV2}: Inverted residuals and linear
  bottlenecks}, in: \bibinfo{booktitle}{Proceedings of the IEEE Conference on
  Computer Vision and Pattern Recognition}, pp. \bibinfo{pages}{4510--4520}.
\bibitem[{Selvaraju et~al.(2017)Selvaraju, Cogswell, Das, Vedantam, Parikh and
  Batra}]{bib:selvaraju2017grad}
\bibinfo{author}{Selvaraju, R.R.}, \bibinfo{author}{Cogswell, M.},
  \bibinfo{author}{Das, A.}, \bibinfo{author}{Vedantam, R.},
  \bibinfo{author}{Parikh, D.}, \bibinfo{author}{Batra, D.},
  \bibinfo{year}{2017}.
\newblock \bibinfo{title}{{GRAD-CAM}: Visual explanations from deep networks
  via gradient-based localization}, in: \bibinfo{booktitle}{Proceedings of the
  IEEE International Conference on Computer Vision}, pp.
  \bibinfo{pages}{618--626}.
\bibitem[{Shiraishi et~al.(2000)Shiraishi, Katsuragawa, Ikezoe, Matsumoto,
  Kobayashi, Komatsu, Matsui, Fujita, Kodera and
  Doi}]{bib:shiraishi2000development}
\bibinfo{author}{Shiraishi, J.}, \bibinfo{author}{Katsuragawa, S.},
  \bibinfo{author}{Ikezoe, J.}, \bibinfo{author}{Matsumoto, T.},
  \bibinfo{author}{Kobayashi, T.}, \bibinfo{author}{Komatsu, K.i.},
  \bibinfo{author}{Matsui, M.}, \bibinfo{author}{Fujita, H.},
  \bibinfo{author}{Kodera, Y.}, \bibinfo{author}{Doi, K.},
  \bibinfo{year}{2000}.
\newblock \bibinfo{title}{Development of a digital image database for chest
  radiographs with and without a lung nodule: receiver operating characteristic
  analysis of radiologists' detection of pulmonary nodules}.
\newblock \bibinfo{journal}{American Journal of Roentgenology}
  \bibinfo{volume}{174}, \bibinfo{pages}{71--74}.
\bibitem[{Shrikumar et~al.(2017)Shrikumar, Greenside and
  Kundaje}]{bib:shrikumar2017learning}
\bibinfo{author}{Shrikumar, A.}, \bibinfo{author}{Greenside, P.},
  \bibinfo{author}{Kundaje, A.}, \bibinfo{year}{2017}.
\newblock \bibinfo{title}{Learning important features through propagating
  activation differences}, in: \bibinfo{booktitle}{International Conference on
  Machine Learning}, \bibinfo{organization}{PMLR}. pp.
  \bibinfo{pages}{3145--3153}.
\bibitem[{Signoroni et~al.(2021)Signoroni, Savardi, Benini, Adami, Leonardi,
  Gibellini, Vaccher, Ravanelli, Borghesi, Maroldi
  et~al.}]{bib:signoroni2021bs}
\bibinfo{author}{Signoroni, A.}, \bibinfo{author}{Savardi, M.},
  \bibinfo{author}{Benini, S.}, \bibinfo{author}{Adami, N.},
  \bibinfo{author}{Leonardi, R.}, \bibinfo{author}{Gibellini, P.},
  \bibinfo{author}{Vaccher, F.}, \bibinfo{author}{Ravanelli, M.},
  \bibinfo{author}{Borghesi, A.}, \bibinfo{author}{Maroldi, R.}, et~al.,
  \bibinfo{year}{2021}.
\newblock \bibinfo{title}{{BS-Net}: Learning {COVID-19} pneumonia severity on a
  large chest {X-ray} dataset}.
\newblock \bibinfo{journal}{Medical Image Analysis} \bibinfo{volume}{71},
  \bibinfo{pages}{102046}.
\bibitem[{Simonyan and Zisserman(2014)}]{bib:simonyan2014very}
\bibinfo{author}{Simonyan, K.}, \bibinfo{author}{Zisserman, A.},
  \bibinfo{year}{2014}.
\newblock \bibinfo{title}{Very deep convolutional networks for large-scale
  image recognition}.
\newblock \bibinfo{journal}{{arXiv} preprint {arXiv}:1409.1556} .
\bibitem[{Soda et~al.(2021)Soda, D’Amico, Tessadori, Valbusa, Guarrasi,
  Bortolotto, Akbar, Sicilia, Cordelli, Fazzini
  et~al.}]{bib:soda2021aiforcovid}
\bibinfo{author}{Soda, P.}, \bibinfo{author}{D’Amico, N.C.},
  \bibinfo{author}{Tessadori, J.}, \bibinfo{author}{Valbusa, G.},
  \bibinfo{author}{Guarrasi, V.}, \bibinfo{author}{Bortolotto, C.},
  \bibinfo{author}{Akbar, M.U.}, \bibinfo{author}{Sicilia, R.},
  \bibinfo{author}{Cordelli, E.}, \bibinfo{author}{Fazzini, D.}, et~al.,
  \bibinfo{year}{2021}.
\newblock \bibinfo{title}{{AIforCOVID}: predicting the clinical outcomes in
  patients with {COVID-19} applying {AI} to {chest-X-rays}: an italian
  multicentre study}.
\newblock \bibinfo{journal}{Medical Image Analysis} \bibinfo{volume}{74},
  \bibinfo{pages}{102216}.
\bibitem[{Stahlschmidt et~al.(2022)Stahlschmidt, Ulfenborg and
  Synnergren}]{bib:stahlschmidt2022multimodal}
\bibinfo{author}{Stahlschmidt, S.R.}, \bibinfo{author}{Ulfenborg, B.},
  \bibinfo{author}{Synnergren, J.}, \bibinfo{year}{2022}.
\newblock \bibinfo{title}{Multimodal deep learning for biomedical data fusion:
  a review}.
\newblock \bibinfo{journal}{Briefings in Bioinformatics} .
\bibitem[{Sundararajan et~al.(2017)Sundararajan, Taly and
  Yan}]{bib:sundararajan2017axiomatic}
\bibinfo{author}{Sundararajan, M.}, \bibinfo{author}{Taly, A.},
  \bibinfo{author}{Yan, Q.}, \bibinfo{year}{2017}.
\newblock \bibinfo{title}{Axiomatic attribution for deep networks}, in:
  \bibinfo{booktitle}{International Conference on Machine Learning},
  \bibinfo{organization}{PMLR}. pp. \bibinfo{pages}{3319--3328}.
\bibitem[{Szegedy et~al.(2015)Szegedy, Liu, Jia, Sermanet, Reed, Anguelov,
  Erhan, Vanhoucke and Rabinovich}]{bib:szegedy2015going}
\bibinfo{author}{Szegedy, C.}, \bibinfo{author}{Liu, W.}, \bibinfo{author}{Jia,
  Y.}, \bibinfo{author}{Sermanet, P.}, \bibinfo{author}{Reed, S.},
  \bibinfo{author}{Anguelov, D.}, \bibinfo{author}{Erhan, D.},
  \bibinfo{author}{Vanhoucke, V.}, \bibinfo{author}{Rabinovich, A.},
  \bibinfo{year}{2015}.
\newblock \bibinfo{title}{Going deeper with convolutions}, in:
  \bibinfo{booktitle}{Proceedings of the IEEE Conference on Computer Vision and
  Pattern Recognition}, pp. \bibinfo{pages}{1--9}.
\bibitem[{Tan et~al.(2019)Tan, Chen, Pang, Vasudevan, Sandler, Howard and
  Le}]{bib:tan2019mnasnet}
\bibinfo{author}{Tan, M.}, \bibinfo{author}{Chen, B.}, \bibinfo{author}{Pang,
  R.}, \bibinfo{author}{Vasudevan, V.}, \bibinfo{author}{Sandler, M.},
  \bibinfo{author}{Howard, A.}, \bibinfo{author}{Le, Q.V.},
  \bibinfo{year}{2019}.
\newblock \bibinfo{title}{{MnasNet}: Platform-aware neural architecture search
  for mobile}, in: \bibinfo{booktitle}{Proceedings of the IEEE/CVF Conference
  on Computer Vision and Pattern Recognition}, pp. \bibinfo{pages}{2820--2828}.
\bibitem[{Wu and Arribas(2003)}]{bib:wu2003fusing}
\bibinfo{author}{Wu, Y.}, \bibinfo{author}{Arribas, J.I.},
  \bibinfo{year}{2003}.
\newblock \bibinfo{title}{Fusing output information in neural networks:
  Ensemble performs better}, in: \bibinfo{booktitle}{Proceedings of the 25th
  Annual International Conference of the IEEE Engineering in Medicine and
  Biology Society}, \bibinfo{organization}{IEEE}. pp.
  \bibinfo{pages}{2265--2268}.
\bibitem[{Wynants et~al.(2020)Wynants, Van~Calster, Collins, Riley, Heinze,
  Schuit, Bonten, Dahly, Damen, Debray et~al.}]{bib:wynants2020prediction}
\bibinfo{author}{Wynants, L.}, \bibinfo{author}{Van~Calster, B.},
  \bibinfo{author}{Collins, G.S.}, \bibinfo{author}{Riley, R.D.},
  \bibinfo{author}{Heinze, G.}, \bibinfo{author}{Schuit, E.},
  \bibinfo{author}{Bonten, M.M.}, \bibinfo{author}{Dahly, D.L.},
  \bibinfo{author}{Damen, J.A.}, \bibinfo{author}{Debray, T.P.}, et~al.,
  \bibinfo{year}{2020}.
\newblock \bibinfo{title}{Prediction models for diagnosis and prognosis of
  {COVID-19}: systematic review and critical appraisal}.
\newblock \bibinfo{journal}{BMJ} \bibinfo{volume}{369}.
\bibitem[{Zadeh et~al.(2017)Zadeh, Chen, Poria, Cambria and
  Morency}]{bib:zadeh2017tensor}
\bibinfo{author}{Zadeh, A.}, \bibinfo{author}{Chen, M.},
  \bibinfo{author}{Poria, S.}, \bibinfo{author}{Cambria, E.},
  \bibinfo{author}{Morency, L.P.}, \bibinfo{year}{2017}.
\newblock \bibinfo{title}{Tensor fusion network for multimodal sentiment
  analysis}.
\newblock \bibinfo{journal}{arXiv preprint arXiv:1707.07250} .
\bibitem[{Zhu et~al.(2020)Zhu, Ge, Jiang, Zhang, Li, Zhao, Zhang and
  Duong}]{bib:zhu2020deep}
\bibinfo{author}{Zhu, J.S.}, \bibinfo{author}{Ge, P.}, \bibinfo{author}{Jiang,
  C.}, \bibinfo{author}{Zhang, Y.}, \bibinfo{author}{Li, X.},
  \bibinfo{author}{Zhao, Z.}, \bibinfo{author}{Zhang, L.},
  \bibinfo{author}{Duong, T.Q.}, \bibinfo{year}{2020}.
\newblock \bibinfo{title}{Deep-learning artificial intelligence analysis of
  clinical variables predicts mortality in {COVID-19} patients}.
\newblock \bibinfo{journal}{Journal of the American College of Emergency
  Physicians Open} \bibinfo{volume}{1}, \bibinfo{pages}{1364--1373}.

\end{thebibliography}
\end{document}